\documentclass[aps,physrev,twocolumn, nofootinbib]{revtex4-2}

\usepackage{hyperref}

\usepackage{amsmath}
\usepackage{comment}
\usepackage{booktabs} 
\usepackage{graphicx}
\usepackage{xcolor}
\usepackage[T1]{fontenc}

\usepackage{newtxtext,newtxmath}
\usepackage{mathtools}
\usepackage{calc}
\newcommand{\mathtextover}[3][l]{\mathmakebox[\widthof{\ensuremath{\displaystyle #3}}][#1]{\ensuremath{\displaystyle #2}}}

\newcommand{\aeq}{\ensuremath{a_\mathrm{eq}}}

\newcommand{\etaEM}{\ensuremath{\eta_\mathrm{EM}}}

\DeclareMathOperator{\sign}{sign}

\begin{document}

\title{Evidence for Low Universal Equilibrium Black Hole Spin in Luminous Magnetically Arrested Disks}

\author{Beverly Lowell}
\email{beverlylowell@u.northwestern.edu}
\affiliation{Center for Interdisciplinary Exploration $\&$ Research in Astrophysics (CIERA), Physics and Astronomy, Northwestern University, Evanston, IL 60201, USA}

\author{Jonatan Jacquemin-Ide}
\affiliation{JILA, University of Colorado and National Institute of Standards and Technology, 440 UCB, Boulder, CO 80309-0440, USA}
\affiliation{Center for Interdisciplinary Exploration $\&$ Research in Astrophysics (CIERA), Physics and Astronomy, Northwestern University, Evanston, IL 60201, USA}

\author{Matthew Liska}

\affiliation{Center for Relativistic Astrophysics, Georgia Institute of Technology, Howey Physics Bldg, 837 State St NW, Atlanta, GA, 30332, USA}

\author{Alexander Tchekhovskoy}
\affiliation{Center for Interdisciplinary Exploration $\&$ Research in Astrophysics (CIERA), Physics and Astronomy, Northwestern University, Evanston, IL 60201, USA}
\affiliation{NSF-Simons AI Institute for the Sky (SkAI), 172 E. Chestnut St., Chicago, IL 60611, USA}

\date{\today}

\begin{abstract}
Relativistic collimated outflows, or jets, provide a crucial mode of active galactic nucleus feedback. Although the jets extract their energy from the black hole (BH) rotation, their effect on the BH spin is poorly understood. Because the spin controls radiative and mechanical BH feedback, lack of first-principles models for BH spin evolution limits our ability to interpret observations, including the recent LIGO-Virgo-KAGRA spin constraints. Particularly important are luminous disks, which rapidly grow and strongly torque their BHs. Jetless and weakly magnetized ``standard'' luminous disks \emph{spin up} their BHs to near-maximum dimensionless spin, $a_\mathrm{eq,NT}=0.998$. However, sufficient large-scale vertical magnetic flux can cause the inner disk to enter a magnetically arrested disk (MAD) state, whose jets can efficiently extract BH rotational energy and significantly \emph{spin down} the BH. Indeed, \citet{lowell_rapid_2023} found that nonradiative, thick MADs \emph{spin down} their BHs to very low $a_\mathrm{eq,MAD}^\mathrm{thick}=0.07$. Moreover, their analytic model predicted that luminous, thin MADs also \emph{spin down} their BHs to low $a_\mathrm{eq,MAD}^\mathrm{thin}\sim0.3\text{--}0.5$. To test this prediction, we perform 3D general relativistic (radiation) magnetohydrodynamic (GR(R)MHD) simulations of MADs across a wide range of BH spin ($-0.9\le{}a\le0.99$) and disk dimensionless thickness ($0.03\le{}h/r\le0.3$, which corresponds to Eddington ratio, $0.35\le{}\dot{m}/\dot{m}_\mathrm{Edd}\le\infty$). We find that luminous, thin MADs ($0.03\le{}h/r\le0.1$) efficiently \emph{spin down} their BHs to a low \emph{universal} equilibrium spin value, $a_\mathrm{eq,MAD}^\mathrm{thin}\approx0.3$: a maximally spinning BH ($a=1$) spins down to $a=0.5$ after accreting just $25\%$ of its initial mass. Our results are consistent with quadratic convergence, $a_\mathrm{eq,MAD}^\mathrm{fit}\simeq0.3-2.7(h/r)^2\to0.3$ as $h/r\to0$, which we attribute to the aggressive cooling that renders disk thermodynamics irrelevant and magnetic forces insensitive to thermal $h/r$.
We finish by discussing the astrophysical implications.

\end{abstract}

\maketitle

\section{Introduction} \label{sec:intro}

Relativistic jets are observed across a range of astrophysical objects, including active galactic nuclei (AGN), ranging from low luminosity AGN (e.g., M87) to luminous quasars \citep{kong_black_2018}, x-ray binaries \citep[XRBs,][]{done_modelling_2007}, tidal disruption events \citep[TDEs,][]{burrows_relativistic_2011,bloom_possible_2011,tchekhovskoy_swift_2014}, and gamma ray bursts (GRBs) powered by collapsars \citep{woosley_gamma-ray_1993,goldstein_estimating_2016} and neutron star (NS) mergers \citep{foucart_post-merger_2015,metzger_kilonovae_2019, nakar_electromagnetic_2020}.

A popular mechanism for jet launching is the Blandford-Znajek (BZ) process \citep{blandford_electromagnetic_1977}: large-scale vertical magnetic flux taps into the spin energy of the black hole (BH) and leads to BH spin-down. The jet power scales approximately as $P_\mathrm{jet} \propto \Phi^2_\mathrm{BH} a^2$, where $\Phi_\mathrm{BH}$ is the vertical magnetic flux on the BH, and $-1 \le a \le 1$ is the dimensionless BH spin \citep{blandford_electromagnetic_1977,1999MNRAS.308.1069K,tchekhovskoy_black_2010}. 

Because BH spin controls jet power, it is crucial to 
understand what determines the spin across different astrophysical sources. However, observationally constraining the BH spin is challenging and often subject to tension, as different measurement methods yield incompatible spin distributions. Reflection methods in XRBs and AGN consistently measure moderate to high BH spins, $a\gtrsim0.5$ \citep{reynolds_observational_2021,draghis_systematically_2024}. In contrast, gravitational wave observations by LIGO-Virgo-KAGRA (LVK) consistently prefer small spins, $a\lesssim 0.4$, from waveform data \citep{abbott_binary_2019,wysocki_reconstructing_2019,ligo_scientific_collaboration_population_2023,edelman_cover_2023}. Constraining and understanding spin evolution is crucial for revealing the formation pathways of LVK sources and advancing our knowledge of binary evolution, because up to $\sim50\%$ of LVK sources can come from stellar binary systems 
\citep{edelman_cover_2023}.
 Accretion flows can directly affect the BH spin: the accreted material adds angular momentum to the BH and leads to its spin-up. Indeed, in the Novikov-Thorne (NT) accretion disk model \citep{novikov_astrophysics_1973}, which does not include large-scale magnetic fields, the BH spins up due to accreting (Keplerian) disk's angular momentum \citep{bardeen_kerr_1970,moderski_black_1996}.

Due to the no-hair theorem, BHs cannot hold onto the magnetic flux by themselves. Matter in the form of an accretion disk needs to force the magnetic flux into the BH and keep it there. Thus, in jetted accretion systems, where a large scale vertical magnetic field is present, spin evolution will depend on the interplay of accretion and ejection processes near the BH event horizon. Jet shape determines the jet angular momentum and power fluxes and thus the spin evolution of the BH \citep{blandford_electromagnetic_1977, tchekhovskoy_black_2010}. Since the disk dimensionless thickness, $h/r$, or the ratio of disk height to radius, is connected to the jet opening angle, the diet of the BH affects spin evolution. Thus, the BH spin extraction can depend on the disk geometric properties, which in turn are determined by the disk cooling regime, as we now discuss. 

BH accretion rate $\dot m$ and how close it is to the Eddington value $\dot m_\mathrm{Edd}$, at which the radiation becomes as important as gravity, roughly sets the disk cooling regime \citep[e.g.,][]{2014ARA&A..52..529Y}. In the highly super-Eddington regime ($\dot m\gg \dot m_\mathrm{Edd}$), the disk is radiatively inefficient: photons are trapped and advected with the gas \citep{popham_hyperaccreting_1999,wyithe_photon_2012,sadowski_photon-conserving_2015,inayoshi_hyper-eddington_2016}. 
This results in a hot accretion flow, which takes the form of a geometrically thick accretion disk. In numerical simulations, such disks typically have a thermal scale height of $h/r \simeq 0.3-0.5$ \citep{chatterjee_flux_2022}. 
The extremely sub-Eddington regime, $\dot m\ll 0.01 \dot m_\mathrm{Edd}$, also leads to a thick radiatively inefficient accretion flow (RIAF), which takes the form of a thick disk \citep{1994ApJ...428L..13N, 1995ApJ...444..231N, 1995ApJ...452..710N}. 
In the middle, $0.01\dot m_\mathrm{Edd} \lesssim \dot m \lesssim \dot m_\mathrm{Edd}$, lies the moderately sub-Eddington regime with an accretion disk that radiates efficiently and cools to scale height values that can become very small, $h/r\ll1$ \citep{shakura_black_1973,narayan_advection-dominated_1995}. 

In XRBs, the Eddington ratio of the BH correlates with the accretion spectral state \citep{esin_advection-dominated_1997,esin_spectral_1998, qiao_model_2013,gilfanov_x-ray_2010}.
The soft state, with a predominantly thermal spectrum, is associated with razor-thin disks, whose inner edge is at (or close to) the innermost stable circular orbit \citep[ISCO,][]{pringle_accretion_1972, shakura_black_1973,mitsuda_energy_1984,marcel_unified_2019}. The low/hard state, with a nonthermal power-law spectrum, is linked to thicker disks in the extremely sub-Eddington regime, with electrons that are much cooler than the ions \citep{marcel_unified_2018,marcel_unified_2019,liska_formation_2022}. 

AGN also come in a wide range of Eddington ratios. Generally, we expect luminous disks to be geometrically thinner than radiatively inefficient disks, although the iron opacity bump and magnetic support may complicate the picture \citep[e.g.,][]{jiang_opacity-driven_2020,2024arXiv240816856J,2024OJAp....7E..19H,2024arXiv240905467S,2024arXiv241001877K}. About $10 \%$ of luminous quasars are radio-loud, which seems to be related to the strength of their jets \citep{kellermann_vla_1989,ivezic_optical_2002,kellermann_radio-loud_2016}. Like XRBs, the measured AGN BH spins are often large \citep{reynolds_observational_2021}.

More extreme accretion systems, such as collapsing massive stars producing BHs (collapsars) and binary compact object mergers, can achieve ultra-high super-Eddington accretion rates at which neutrino cooling can reduce their aspect ratio to $h/r\sim0.2$  \citep{2007ApJ...657..383C}.

This variety of disk thickness values across different types of astrophysical sources motivates a systematic exploration of the effects that the disk thickness has on the BH spin evolution. For this, it is essential to account for gas accretion and ejection physics, including the effects of strong magnetic field, radiation, and curved spacetime. This ambitious goal can be achieved with general relativistic (radiation) magnetohydrodynamic (GR(R)MHD) simulations. In this work, we focus on magnetically arrested disks (MADs), which are the natural end state of a disk with a sufficiently large vertical magnetic flux reservoir \citep{tchekhovskoy_efficient_2011,2015MNRAS.447..327T,jacquemin-ide_magnetic_2021}. Such highly magnetized disk states have also been found to emerge naturally due to dynamo action within the accretion disk \citep{liska_large-scale_2020,jacquemin-ide_magnetorotational_2024}. MADs can also form in transient sources whose mass accretion rate decreases over time, e.g., TDEs \citep{tchekhovskoy_swift_2014,kelley_tidal_2014} and compact binary mergers \citep{fernandez_mass_2018,christie_role_2019,gottlieb_large-scale_2023}.

BH equilibrium spin is defined as the BH spin for which the angular momentum lost through the jet balances the angular momentum and mass supplied by the disk. For thick, nonradiative MADs, GRMHD simulations revealed that the equilibrium spin is small, $a_\mathrm{eq,MAD}^\mathrm{thick}\simeq 0.035\text{--}0.07$ \citep{2012JPhCS.372a2040T,2015ASSL..414...45T,narayan_jets_2022,lowell_rapid_2023}. \citet{lowell_rapid_2023} found that the BH only needs to accrete $20\%$ of its initial mass to spin down from $a=1$ to $a=0.2$. They also found that the BH loses most of its spin energy to the jets and that the positive angular momentum contribution from the accretion flow is suppressed by the large-scale electromagnetic torques in the accretion disk (see also \citep{2024ApJ...965..175M}). They then used this insight to derive a semi-analytic model for spin evolution in the thick nonradiative MAD regime. Using this model, they predicted that luminous, thin MADs would have low equilibrium spin values, $\aeq \sim 0.3\text{--}0.5$. This is much lower than what is obtained by \citet{ricarte_recipes_2023} based on their GRMHD simulations ($\aeq\sim0.8$), expected in NT disks ($\aeq\sim1$), and measured in weakly magnetized ``standard and normal evolution'' (SANE) accretion disks ($\aeq\sim 0.9$) \citep{gammie_black_2004,chatterjee_flux_2022}.

In this work, we perform high-resolution 3D GRMHD simulations of cooled BH accretion disks in the MAD state to study BH spin evolution for thin accretion disks. This allows us to extend the work of \citet{lowell_rapid_2023} from the radiatively inefficient thick to luminous MADs of varying disk scale height to physically interpret our simulation results.  

Section~\ref{subsec:spin_eqns} presents the BH spin evolution equations. Section \ref{subsec:NTdisk} discusses the standard Novikov-Thorne disk model~\citep{novikov_astrophysics_1973}. Section~\ref{subsec:simulations} describes our numerical setup and simulations. Section~\ref{subsec:calc_fluxes} explains how we calculate the torques on the BH. Section~\ref{sec:results_spinevolution} lays out our key results on spin evolution in thin MADs. Section~\ref{sec:coolmodel} derives our new thin MAD spin evolution model and spin-down timescales. Section~\ref{sec:jet_struc_force} explores the dynamic link between the jet and disk magnetic structures, and their effect on the spin-down. Finally, Section~\ref{sec:discussion} discusses the results and concludes. Throughout, we use units such that $G = c = 1$. We adopt Lorentz-Heaviside units for the magnetic field, so that we absorb the factor of $(4\pi)^{-1/2}$ into the definition of the magnetic field. As a result, the magnetic pressure is $p_\mathrm{mag} = b^2/2$, where  $b$ is the comoving magnetic field strength.

\section{Methods and Equations} \label{sec:methods}

\subsection{Spin evolution equations} \label{subsec:spin_eqns}
We define the dimensionless BH spin as 
\begin{equation}
    a = \frac{J}{M^2},
\end{equation}
where $J$ is the angular momentum  and $M$ is the mass of the BH, so that it spans $-1 \leq a \leq 1$. We can then compute the spin evolution of the BH by solving a set of coupled ODEs for the total energy,
\begin{align}
    dM&= e_\mathrm{in} dm
    \label{eq:e_ODE}\\
\intertext{and angular momentum,}
    dJ &= l_\mathrm{in} dm,
    \label{eq:l_ODE}
\end{align}
accreted by the BH \citep{moderski_black_1996,lowell_rapid_2023}, where $l_\mathrm{in}$ and $e_\mathrm{in}$ are the specific angular momentum and energy fluxes, respectively, on the BH horizon,  $r_\mathrm{H} = r_\mathrm{g}(1 + \sqrt{1-a^2})$, and $dm$ is the amount of accreted mass. Here, $r_\mathrm{g} = GM/c^2$ is the BH gravitational radius.
Combining Equations~\eqref{eq:e_ODE} and \eqref{eq:l_ODE} gives us the ODE for the change in BH spin with BH mass,
\begin{equation}
    \frac{da}{d \ln M} = \frac{l_\mathrm{in}}{e_\mathrm{in}} - 2a = \frac{s}{e_\mathrm{in}},
    \label{eq:da_dlogM}
\end{equation} 
where $\ln$ is the natural logarithm, and we have defined the dimensionless spin-up parameter,
\begin{equation}
    s \equiv \frac{da}{dm} M = l_\mathrm{in} - 2 a e_\mathrm{in},
    \label{eq:s_param}
\end{equation}
which roughly tells us the change in spin after the BH accretes its own mass (i.e., $dm \sim M$).
Computing the BH spin evolution is a question of determining, as a function of BH spin, the specific fluxes, $l_\mathrm{in}$ and $e_\mathrm{in}$, both of which depend on the (to be determined) accretion flow and jet structures. 

\subsection{Novikov-Thorne disk model} \label{subsec:NTdisk}

The NT disk  model \citep{novikov_astrophysics_1973} is an analytic model for a razor-thin accretion disk where the orbits are Keplerian, and the accretion mechanism is a viscosity of turbulent origin. The specific energy and angular momentum fluxes measured at the BH event horizon are  \citep{bardeen_kerr_1970,moderski_black_1996},
\begin{align}
   e_\mathrm{in} &= \left(1-\frac{2 r_\mathrm{g}}{3 R_\mathrm{ISCO}}\right)^{1/2}\\
\intertext{and}
   l_\mathrm{in} &= \frac{2M}{3^{3/2}} \left[1 + 2(3R_\mathrm{ISCO}/r_\mathrm{g} - 2)^{1/2}\right],
\end{align}
where $R_\mathrm{ISCO}$ is the radius of the innermost stable circular orbit \citep[ISCO,][]{1983bhwd.book.....S},
\begin{align}
  R_\mathrm{ISCO}/r_\mathrm{g} &= 3 + Z_2 - \sign (a) \left[(3-Z_1)(3+Z_1+2Z_2)\right]^{1/2},\\
\intertext{where}
    Z_1 &= 1+(1-a^2)^{1/3} \left[(1+a)^{1/3} + (1-a)^{1/3}\right],\\
   Z_2 &= (3a^2 + Z_1^2)^{1/2}.
\end{align}

\subsection{Suite of MAD simulations} \label{subsec:simulations}

We carry out simulations of MADs to compute the energy and angular momentum fluxes into the BH. For this, we use \texttt{H-AMR} \citep{2022ApJS..263...26L}, a 3D massively parallel graphical processing unit (GPU) accelerated GRRMHD code, which is based on 3D \texttt{HAMRPI} \citep{2015MNRAS.454.1848R,2017MNRAS.467.3604R,2019ascl.soft12014T} and 2D \texttt{HARM2D}  \citep{gammie_harm_2003, noble_primitive_2006} codes. We initialize the simulations with a torus in hydrostatic equilibrium \citep{fishbone_relativistic_1976} about a BH, in a Kerr-Schild metric (see Sec.~\ref{sec:nonrad-mads}--\ref{sec:rad-mads}). We adopt spherical polar coordinates, $r$, $\theta$, $\varphi$, and use a uniform grid in $\ln r$, $\theta$, and $\varphi$. The radial grid extends from just inside the event horizon, $R_\mathrm{in} = 0.8r_\mathrm{H}$, to $R_\mathrm{out} = 10^5r_\mathrm{g}$ (or $R_\mathrm{out} = 10^4r_\mathrm{g}$). 
All our simulations have at least five radial cells inside of the event horizon; this ensures that the exterior of the BH is causally disconnected from the inner radial boundary of the grid. We adopt outflow boundary conditions (BCs) at the inner and outer radial boundaries, transmissive BCs at the polar boundaries, and periodic BCs in the $\varphi$-direction \citep{2022ApJS..263...26L}.

We insert a large poloidal (i.e., with magnetic field pointing in the $R$- and $z$-directions) magnetic flux loop into the torus, with the covariant magnetic vector potential described below.  We wait for the large-scale disk magnetic flux to flood the BH and lead to the MAD state \citep[e.g.,][]{tchekhovskoy_efficient_2011}. 

To compute the evolution of the BH spin over cosmological timescales, which are far longer than the duration of any of our simulations, we perform a suite of GRMHD simulations spanning a wide range of BH spin values. We keep the BH spin constant throughout each of the simulations, a good approximation given the extremely short duration of our simulations in the cosmological context. However, in each of the simulations, we compute the torques acting on the BH. We then use these to model the BH spin evolution by solving a set of coupled ODEs. To understand how the disk thickness affects the BH spin evolution, we have run the simulations with dimensionless thermal disk scale heights ranging from thick, $h/r=0.3$, to thinner, $h/r = 0.1$ and thinnest, $h/r = 0.03$, disk. 

\begin{figure*}[!tbp]
    \includegraphics[width=0.7\textwidth]{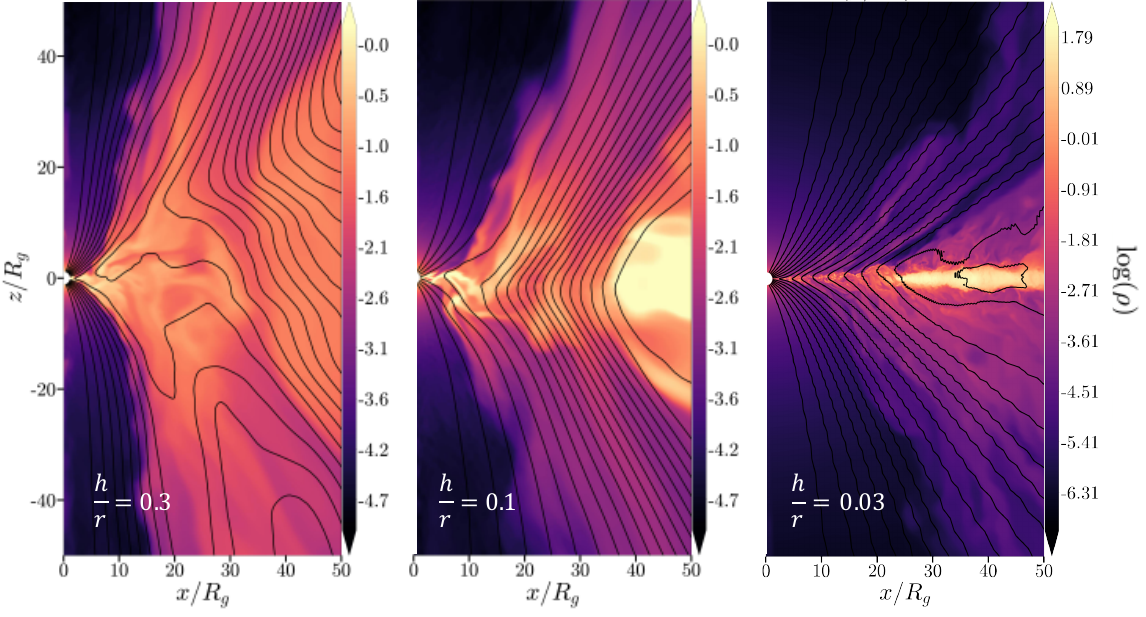}%
  \caption{Smaller $h/r$ MADs (yellow) are thinner and result in wider, less collimated polar jets (dark purple), as seen in the vertical slices through the instantaneous fluid-frame density for our simulations with different disk scale heights, $h/r = 0.3$ (model H3a0.94hr), $0.1$ (model H1a0.94), and $0.03$ (model Ra0.94), as labeled on the panels. Black lines show the axisymmetric poloidal magnetic flux contours and demonstrate that the vertical magnetic flux floods both the BH and inner disk, as expected in MADs. }
  \label{fig:density_maps}
\end{figure*}

Figure~\ref{fig:density_maps} shows the vertical slices through density in our simulations of MADs around a rapidly spinning BH, $a=0.9375$, for these three disk thickness values. We see that the smaller $h/r$ disks are thinner and result in wider jets.
Whereas most our simulations use a radiative cooling prescription, for which $h=\sqrt{2/\pi}c_\mathrm{i}/\Omega$, where $c_\mathrm{i}$ is the isothermal sound speed and $\Omega$ is the angular frequency of the gas \citep{noble_direct_2009}, some use the full radiation transport scheme with a two-moment (M1) radiative closure~\citep{liska_formation_2022}.
We summarize the main simulation parameters in Table~\ref{tab:sim_dets_hovr1} and discuss the simulations below in more detail.

\begin{table*}[!htp]

\centering
\caption{Simulation details. Column 1: model name. Column 2: BH spin. Column 3: thermal disk scale height. Column 4: polytropic index. Column 5: spin-up parameter. Column 6: azimuthal wedge. Column 7: base grid resolution. Column 8: effective resolution within the disk. Column 9: simulation duration. Column 10: time-averaging window. 
}
\begin{tabular}{
  l
  c
  c
  c
  c
  c
  c
  c
  c
  c
  @{}
}
\toprule
 & $a$ & $h/r$ & $\Gamma$ & $s$ & $\Delta_\varphi$ & ($N_r \times N_\theta \times N_\varphi$)$_\mathrm{base}$ & ($N_r \times N_\theta \times N_\varphi$)$_\mathrm{eff}$ & $t_\mathrm{run}\, [r_\mathrm{g}/c]$ & $t_\mathrm{avg}\, [r_\mathrm{g}/c]$ \\
\hline
\hline
\multicolumn{10}{@{}c}{$\mathit{h/r = 0.3}$, $\mathit{\Gamma = 4/3}$} \\

         H3a-0.9 & $-0.9$ & $0.3$  & $4/3$ & 5.98 & $2\pi$ &  $288\times 128\times 64$ & $288\times 224\times 64$  & ($0; 20100$) & ($10000; 20100$) \\
         H3a-0.5 & $-0.5$ & $0.3$  & $4/3$ & 5.33 & $\pi$  & $288\times 128\times 32$ & $288\times 224\times 32$ & ($0; 16350$) & ($10000; 16350$) \\
         H3a-0.2 & $-0.2$ & $0.3$  & $4/3$ & 3.61 & $\pi$ &  $288\times 128\times 32$ & $288\times 224\times 32$ & ($0; 15200$) & ($10000; 15200$) \\
         H3a0.0  & $0.0$ & $0.3$  & $4/3$ & 1.03 & $\pi$ &  $288\times 128\times 32$ & $288\times 224\times 32$ &($0; 18550$) & ($10000; 18550$)  \\
         H3a0.1  & $0.1$ & $0.3$  & $4/3$ & -0.5 & $\pi$ &  $288\times 128\times 32$ & $288\times 224\times 32$ &($0; 18725$) & ($10000; 18725$)  \\
         H3a0.2  & $0.2$ & $0.3$  & $4/3$ & -1.56 & $\pi$ &  $288\times 128\times 32$ & $288\times 224\times 32$ & ($0; 13400$) & ($10000; 13400$) \\
         H3a0.5  & $0.5$ & $0.3$  & $4/3$ &  -4.62 & $\pi$ &  $288\times 128\times 32$ & $288\times 224\times 32$ & ($0; 13050$) & ($10000; 13050$)  \\
         H3a0.9  & $0.9$ & $0.3$  & $4/3$ & -7.47 & $2\pi$ &  $288\times 128\times 64$ & $288\times 224\times 64$ & ($0; 19900$) & ($10000; 19900$)  \\
         H3a0.99 & $0.99$ & $0.3$  & $4/3$ & -7.39 &  $2\pi$&  $288\times 128\times 64$ & $288\times 224\times 64$ & \phantom{$^\dagger$}($0; 14650$)$^\dagger$   & ($10000; 14650$) \\

\hline
\hline
\multicolumn{10}{@{}c}{$\mathit{h/r = 0.3}$ $HR$, $\mathit{\Gamma = 13/9}$} \\

H3a-0.9hr  &  -0.9 & $0.3$ & $13/9$ & $6.05$ & $2\pi$ & $384\times 300\times 64$ & $384\times 300\times 64$ &  ($0; 20225$)  & ($10000; 20000$)\\
         H3a-0.5hr  &  -0.5 & $0.3$ & $13/9$ & $4.52$ & $2\pi$ & $384\times 300\times 64$ & $384\times 300\times 64$ &  ($0; 23250$)  & ($10000; 23250$)\\
         H3a0.2hr &  0.2 & $0.3$ & $13/9$ & $-2.40$ & $2\pi$ & $384\times 300\times 64$ & $384\times 300\times 64$ &  ($0; 14800$)  & ($10000; 14750$)\\
         H3a0.5hr  &  0.5 & $0.3$ & $13/9$ & $-5.31$ & $2\pi$ & $384\times 300\times 64$ & $384\times 300\times 64$ &  ($0; 57825$) & ($52800; 57800$)\\
         H3a0.94hr  &  0.9375 & $0.3$ & $13/9$ & $-8.67$ & $2\pi$ & $384\times 300\times 64$ & $384\times 300\times 64$ &  ($0; 68105$) & ($63000; 68000$)\\
\hline
\hline
\multicolumn{10}{@{}c}{$\mathit{h/r = 0.1}$} \\
H1a-0.9  &  -0.9 & $0.1$ & $13/9$ & $4.53$ & $2\pi$ & $384\times 300\times 64$ & $384\times 300\times 64$ &  ($0; 65055$) & ($60000; 65000$)\\
         H1a-0.5 & -0.5 &  $0.1$ & $13/9$ & $3.17$ & $2\pi$ & $384\times 300\times 64$ & $384\times 300\times 64$ &  ($0; 60064$) & ($55000; 60000$)\\
         H1a-0.2 & -0.2 &  $0.1$ & $13/9$ & $2.31$ &  $2\pi$ &$384\times 300\times 64$ & $384\times 300\times 64$ & ($0; 61065$) & ($56600; 61600$)       \\
         H1a0.0 & 0.0  &  $0.1$ & $13/9$ & $1.46$ & $2\pi$ & $384\times 300\times 64$ & $384\times 300\times 64$ & ($0; 60220$) & ($55200; 60200$)      \\
         H1a0.2 & 0.2  & $0.1$ &  $13/9$ & $0.57$ & $2\pi$ & $384\times 300\times 64$ & $384\times 300\times 64$ &  ($0; 62400$) & ($57400; 62400$)       \\ 
         H1a0.3 & 0.3  & $0.1$ &  $13/9$ & $-0.09$ & $2\pi$ & $384\times 300\times 64$ & $384\times 300\times 64$ &  ($0; 58800$) & ($53800; 58800$)       \\
         H1a0.5 & 0.5  & $0.1$ &  $13/9$ & $-1.36$ & $2\pi$ & $384\times 300\times 64$ & $384\times 300\times 64$ & ($0; 65600$) & ($60600; 65600$)    \\
         H1a0.94 & $0.9375$ & $0.1$ & $13/9$ & $-3.02$ & $2\pi$ & $384\times 300\times 64$ & $384\times 300\times 64$ & ($0; 56300$) & ($51300; 66300$)    \\
\hline
\hline
\multicolumn{10}{@{}c}{$\mathit{h/r = 0.1}$ $HR$} \\
H1a0.3hr & 0.3  & $0.1$ & $13/9$ & $-0.17$ &  $2\pi$ & $512\times 288\times 256$ & $1024 \times 576\times 512$ &($0; 59400$) & ($45000; 59400$)       \\
\hline
\hline
\multicolumn{10}{@{}c}{$\mathit{h/r = 0.05}$ $LR$   {(warning: unresolved)}} \\

H05a0.1 & 0.1  & $0.05$ &  $13/9$ & $1.51^*$ &  $2\pi$ & $384\times 300\times 64$ & $384\times 300\times 64$ & ($0; 66700$) & ($61700; 66700 $)      \\

H05a0.3 & 0.3  & $0.05$ &  $13/9$ & $0.36^*$ & $2\pi$ & $384\times 300\times 64$ & $384\times 300\times 64$ & ($0; 55800$) & ($50800; 55800$)       \\

H05a0.4 & 0.4  & $0.05$ & $13/9$ & $-0.02^*$ & $2\pi$ & $384\times 300\times 64$ & $384\times 300\times 64$ & ($0; 59380$) & ($54400; 59380$)       \\
\hline
\hline
\multicolumn{10}{@{}c}{$\mathit{h/r = 0.05}$ $HR$} \\

 H05a0.4hr & 0.4  & $0.05$ & $13/9$ & $-0.636$ & $2\pi$ & $512\times 288\times 256$ & $1024\times 576\times 512$ &($0; 45000$) & ($41000; 45000$)       \\

\hline
\hline
\multicolumn{10}{@{}c}{$\mathit{h/r = 0.03}$ $2T$} \\
Ra0.3 & 0.3  & $0.03$ & $5/3$ & $0.04$ & $2\pi$ & $1020 \times 432 \times 288$ & $4080 \times 1728 \times 1152$ &($188839; 193162$) & ($188929; 193162$)    \\
         Ra0.4 & 0.4  & $0.03$ & $5/3$ & -0.39 & $2\pi$ & $1020 \times 432 \times 288$ & $4080 \times 1728 \times 1152$ &($188839; 190869$) & ($188889; 190869$)    \\
         Ra0.94 & $0.9375$ & $0.03$ & $5/3$ & $-3.18$ & $2\pi$ & $1020 \times 432 \times 288$ & $4080 \times 1728 \times 1152$ &($188839; 192879$) & ($188889 ; 192879$)    \\
\hline
\hline
\multicolumn{10}{@{}c}{$\mathit{h/r = 0.03}$ $1T$} \\
H03a0.94 & $0.9375$ & $0.03$ & $5/3$ & $-2.854$ & $2\pi$ & $1020 \times 432 \times 288$ & $4080 \times 1728 \times 1152$ &($188839; 203339$) & ($188839 ; 203339$) \\
\bottomrule
\end{tabular}\\
$^\dagger$ \citep{tchekhovskoy_efficient_2011} continued this run to later time ($t\sim30,500 r_\mathrm{g}/c$) at a higher resolution ($288\times 128\times 128$).\\
* Measurements on unresolved simulations performed for the resolution study described in Appendix \ref{appendix:resolution}.
\label{tab:sim_dets_hovr1}
\end{table*}

\subsubsection{Nonradiative thick MADs}
\label{sec:nonrad-mads}

To study thick MADs we utilize the simulations of \citep{tchekhovskoy_efficient_2011,2012JPhCS.372a2040T,2015ASSL..414...45T} with BH spins $a=-0.9$, $-0.5$, $-0.2$, $0.0$, $0.1$, $0.2$, $0.5$, $0.9$, and $0.99$ and a polytropic index of $\Gamma=4/3$. We label these simulations as $h/r=0.3$ in Table \ref{tab:sim_dets_hovr1}. Collectively, we refer to these simulations as the $h/r=0.3$ simulations. Compared to other simulations in this paper, these simulations are lower resolution (at most $288 \times 128 \times 128$) and shorter duration ($t_\mathrm{final} \le 30,500 r_\mathrm{g}/c$). Here, we have run higher poloidal resolution ($384 \times 300 \times 64$) and longer duration simulations for a range of spin values, $a=-0.9$, $-0.5$, $0.2$, $0.5$, and $0.9375$, with a polytropic index of $\Gamma=13/9$. Collectively, we refer to these simulations as $h/r=0.3\,HR$. We show these higher-resolution thick disk MAD simulations in Table~\ref{tab:sim_dets_hovr1}. Because our thick nonradiative MADs reach a thermal scale height of $h/r \approx 0.3$, we use this fiducial value when referring to these simulations.

\subsubsection{Thin MADs}
\label{sec:cooled-mads}

For two sets of our models, with $h/r = 0.1$ and $h/r = 0.05$, we cool the disk to its target thickness on the Keplerian timescale using a cooling source term following \citep{noble_direct_2009}. For the initial conditions, we adopt an equilibrium hydrodynamic torus \citep{fishbone_relativistic_1976} of inner radius,  $r_\mathrm{in} = 20 r_\mathrm{g}$, and the pressure maximum radius, $r_\mathrm{max} \sim 41 r_\mathrm{g}$, where we tweak the exact value of $r_\mathrm{max}$ such that our torus extends out to an extremely large (but finite) distance, $r_\mathrm{out} \sim \text{few}\times 10^4 r_\mathrm{g}$. 
To quickly achieve the MAD state, we adopt the covariant vector potential, $A_\varphi = q^2$, where (EHT Code Comparison Paper, in preparation),
\begin{equation}
  q = \frac{\rho}{\max\rho} \left(\frac{r}{r_\mathrm{in}}\right)^3 {\sin^3}\theta \, \exp \left( - \frac{r}{400r_\mathrm{g}} \right) - 0.2,
  \label{eq:q}
\end{equation}
which results in a radially extended poloidal magnetic field loop of a sufficiently large poloidal magnetic flux to flood the BH.\footnote{In Eq.~(\ref{eq:q}), the $r^3$ term increases the size and amount of the magnetic flux in the loop by pushing the center of the loop out to larger radii. However, this results in a magnetically dominated outer disk. The exponential term reduces the outer disk magnetization by redistributing some of the magnetic flux back toward smaller radii. The last term in Eq.~(\ref{eq:q}) ensures that the magnetic loop does not extend too close to the torus boundaries.}
We then normalize the magnetic field strength such that the minimum ratio of gas to magnetic pressure is $\min\beta = 100$.

We have carried out both low- and high-resolution simulations for $h/r=0.1$ and $h/r=0.05$. The low-resolution runs use a base grid of $384 \times 300 \times 64$ with no grid refinement, while the high-resolution runs use a base grid of $512 \times 288 \times 256$ with one level of static mesh refinement (SMR) applied for $ |\theta - \pi/2| \leq 0.315 $ and $4\,r_\mathrm{g} \leq r \leq 500\,r_\mathrm{g}$, resulting in an effective resolution of $1024 \times 576 \times 512$ in the disk proper. A resolution convergence study (see Appendix~\ref{appendix:resolution}) confirms that both resolutions are adequate for $h/r=0.1$ simulations. However, thinner disks ($h/r=0.05$) are only resolved at the high-resolution, so we use only the high-resolution $h/r=0.05$ simulations in our analysis.

Additionally, we include a very high-resolution $h/r=0.03$ simulation using a predefined cooling function~\citep{noble_direct_2009}, previously presented in \citep{liska_formation_2022}. We refer to this simulation as model H03a0.94 and also label it as $h/r=0.03$ $1T$. Its resolution matches that of the radiative (GRRMHD) simulations (models RaX) described in Section \ref{sec:rad-mads}. Its initial conditions are described in \citep{liska_formation_2022}.

\subsubsection{Thin radiative MADs}
\label{sec:rad-mads}

Although cooled GRMHD models with fixed $h/r$ offer computational control, they prescribe thermodynamics rather than predict it. GRRMHD models, in contrast, include radiation transport and allow the disk structure to emerge self-consistently, providing the most physically realistic test of whether predefined cooling functions adequately capture the structure of radiatively cooled accretion disks.

To model radiatively cooled MADs with realistic thermodynamics, we use a two-temperature (2T) radiation transport (GRRMHD) simulation of \citep{liska_formation_2022} that results in a disk with a thermal scale height of $h/r \approx 0.03$. This simulation uses the M1 radiation scheme implemented in \texttt{H-AMR}, including the Compton cooling \citep{liska_formation_2022}. To resolve the thin disk structure, the simulation uses a static mesh refinement (SMR) on a base grid of $1020 \times 432 \times 288$ cells. We then quadruple this resolution at $5 r_\mathrm{g} \le r \le 120 r_\mathrm{g}$ and $|\theta-\pi/2| \le 0.13$ to achieve an effective resolution of $4080 \times 1728 \times 1152$ cells in the disk proper. 
The BH spin is $a=0.9375$. We refer to this simulation as model Ra0.94. To improve our BH spin sampling, we restart this model, Ra0.94, at $t=188839 r_\mathrm{g}/c$ for two different spin values, $a=0.3$ (Ra0.3) and $a=0.4$ (Ra0.4). This is the same as the start time of our analysis of Ra0.94 (see Table \ref{tab:sim_dets_hovr1}). We refer to these simulations collectively as $h/r=0.03$ $2T$. Table~\ref{tab:sim_dets_hovr1} summarizes the resolutions and scale heights.

\subsection{Calculating fluxes} \label{subsec:calc_fluxes}

We calculate the angular momentum and energy fluxes by evaluating the components of the stress-energy tensor,
\begin{equation}
    T^{\mu \nu} = T^{\mu \nu}_\mathrm{HD} + T^{\mu \nu}_\mathrm{EM},
    \label{eq:Tmunu}
\end{equation} 
where the hydrodynamic (HD) part is
\begin{equation}
    T^{\mu \nu}_\mathrm{HD} = (\rho + u_\mathrm{g} + p_\mathrm{g})u^\mu u^\nu + p_\mathrm{g} g^{\mu \nu}.
    \label{eq:Tmunu_hydro}
\end{equation} 
Here, $\rho$ is gas density, $u_\mathrm{g}$ is the internal energy, and $p_\mathrm{g}$ is the pressure of the gas, all measured in the fluid frame, $u^\mu$ is the contravariant 4-velocity, and $g^{\mu \nu}$ is the contravariant metric. The electromagnetic (EM) part of the stress-energy tensor is given by 
\begin{equation}
    T^{\mu \nu}_\mathrm{EM} = b^2 u^\mu u^\nu + \frac{1}{2} b^2 g^{\mu \nu} - b^\mu b^\nu
    \label{eq:Tmunu_EM},
\end{equation} 
where $b^\mu$ is the magnetic field four-vector and $b^2 = b^\mu b_\mu = 2 p_\mathrm{mag}$ is twice the magnetic pressure. 

We compute the mass accretion rate, 
\begin{equation}
    \dot{m}(r) = -  \int \langle\rho u^r\rangle dA_{\theta \varphi}, 
        \label{eq:fM}
\end{equation}
and the specific energy and angular momentum fluxes into the BH (defined to be positive when flowing toward the BH),
\begin{align}
    e(r) &= \frac{\dot{E}(r)}{\dot{m}(r)}=\frac{1}{\dot{m}(r)}\int \langle T^r_t \rangle dA_{\theta \varphi},
    \label{eq:fEoverfM}\\
    l(r) &= -\frac{1}{M} \frac{\dot{L}}{\dot{m}(r)}=\frac{1}{M} \frac{{1}}{\dot{m}(r)}\int \langle T^r_\varphi \rangle dA_{\theta \varphi},
    \label{eq:fLoverfM}
\end{align}
where $dA_{\theta \varphi} = \sqrt{-g} \, d\theta \, d\varphi$ is the surface element, and the integrals are taken over a sphere of radius $r$. We define $\langle X \rangle$ as the time average of quantity $X$, computed over the interval, $t_\mathrm{avg}$, specified for each simulation in Table~\ref{tab:sim_dets_hovr1}; more generally, we compute all time averages over $t_\mathrm{avg}$.
When computed at the BH event horizon, the fluxes in eqs.~\eqref{eq:fEoverfM} and~\eqref{eq:fLoverfM} give the prefactors in Equations~\eqref{eq:e_ODE} and~\eqref{eq:l_ODE},  $l_\mathrm{in}\equiv l(r=r_\mathrm{H})$ and $e_\mathrm{in}\equiv e(r=r_\mathrm{H})$. 
Using Eqs.~\eqref{eq:fEoverfM} and \eqref{eq:fLoverfM}, we define $e_\mathrm{HD}$, $e_\mathrm{EM}$, $l_\mathrm{HD}$, and $l_\mathrm{EM}$ by substituting $T^{r}_{t}$ and $T^{r}_{\varphi}$ with their respective hydrodynamic ($T^{\mu\nu}_\mathrm{HD}$) and electromagnetic  ($T^{\mu\nu}_\mathrm{EM}$) components (see definitions in Eqs.~\ref{eq:Tmunu_hydro}, \ref{eq:Tmunu_EM}), where we evaluate all quantities at the event horizon .

We also calculate the direct effect of radiation on the BH using the radiation stress-energy tensor,
\begin{equation}
     R^{\mu \nu} = \frac{4}{3}E_\mathrm{rad}u^{\mu}_\mathrm{rad} u^{\nu}_\mathrm{rad} + \frac{1}{3}E_\mathrm{rad}g^{\mu \nu},
    \label{R_munu} 
\end{equation} 
where $E_\mathrm{rad}$ is the radiation energy density in the radiation frame (which is the frame in which the radiation is isotropic, in the M1 approximation) and $u^{\mu}_\mathrm{rad}$ is the radiation frame 4-velocity. We compute the radiation specific energy and angular momentum fluxes, $e_\mathrm{rad}(r)$ and $l_\mathrm{rad}(r)$, in the same way as in Equations \eqref{eq:fEoverfM} and \eqref{eq:fLoverfM}, respectively, but where we replace $T^{\mu}_{\nu}$ with $R^{\mu}_{\nu}$. We also define the radiation pressure $p_\mathrm{rad}=\frac{1}{3}E_\mathrm{rad}$.

We define the ``jet'' torque on the BH to be due to the electromagnetic components of the angular momentum and energy fluxes on the BH. We define the ``disk'' torque on the BH to be due to their hydrodynamic components. We also define the ``radiation'' torque on the BH due to $e_\mathrm{rad}(r_\mathrm{H})$ and $l_\mathrm{rad}(r_\mathrm{H})$.

Because numerical floors can add nonphysical mass density and internal energy near the horizon, we apply a magnetization cutoff on the disk torque, following the procedure described in Appendix~B of~\citep{lowell_rapid_2023}. We exclude the contribution from the jets by applying a magnetization cut,  $\sigma = b^2/\rho \leq 30$, for models that use the density floor of $\sigma_\mathrm{max}=50$. 
For models with $\sigma_\mathrm{max}=15$, we choose the disk magnetization cut of $\sigma\le10$. We also apply the same $\sigma$-cut to the radiation fluxes, as they are biased by density floors within the jets. Note that although models Ra0.3 and Ra0.94 include this cutoff, model Ra0.4 does not due to data loss. 

We can write the total energy extraction efficiency in terms of the specific energy flux $e_\mathrm{in}$ at the BH,  
\begin{equation}
  \eta = \frac{\dot mc^2 - \dot E}{\dot mc^2} =  1 - e_\mathrm{in} = 1 - e_\mathrm{HD} - e_\mathrm{EM} - e_\mathrm{rad},
  \label{eta_1minusE}
\end{equation}
where for clarity of units we have reinstated the $c^2$ factor. The efficiency, $\eta$, represents the energy return on the rest-mass energy investment into the BH: for an accreted rest-mass energy of $mc^2$, the BH returns $\eta mc^2$ in the mechanical and radiative forms. MADs around rapidly spinning BHs can reach $\eta\sim200$\% \citep{tchekhovskoy_efficient_2011,2012MNRAS.423.3083M,2025arXiv250523888L}. This makes BHs the most efficient energy producers, more than 2 orders of magnitude more efficient than nuclear fusion (which has $\eta \sim0.7$\%). This total efficiency should not be confused with the electromagnetic efficiency, $\eta_{\rm EM}$, defined below.
\section{BH Spin Evolution} \label{sec:results_spinevolution}

Throughout this work, we differentiate simulation families of different disk scale heights by color. Gray points show nonradiative $h/r=0.3$ simulations from \citep{tchekhovskoy_efficient_2011}, previously analyzed in \citep{lowell_rapid_2023}. Blue points show $h/r=0.3$ $HR$ models, purple points show $h/r=0.1$ models, and green points show $h/r=0.05$ $HR$ models. Either red or orange points show M1 2T-radiation simulations, $h/r=0.03$ $2T$. Finally, black points show the $h/r=0.03$ $1T$ simulation, which uses a cooling function \citep{noble_direct_2009}.

\subsection{Spin-up parameter and equilibrium spin} \label{subsec:spinup}
 
\begin{figure}
    \centering
    \includegraphics[width=\columnwidth]{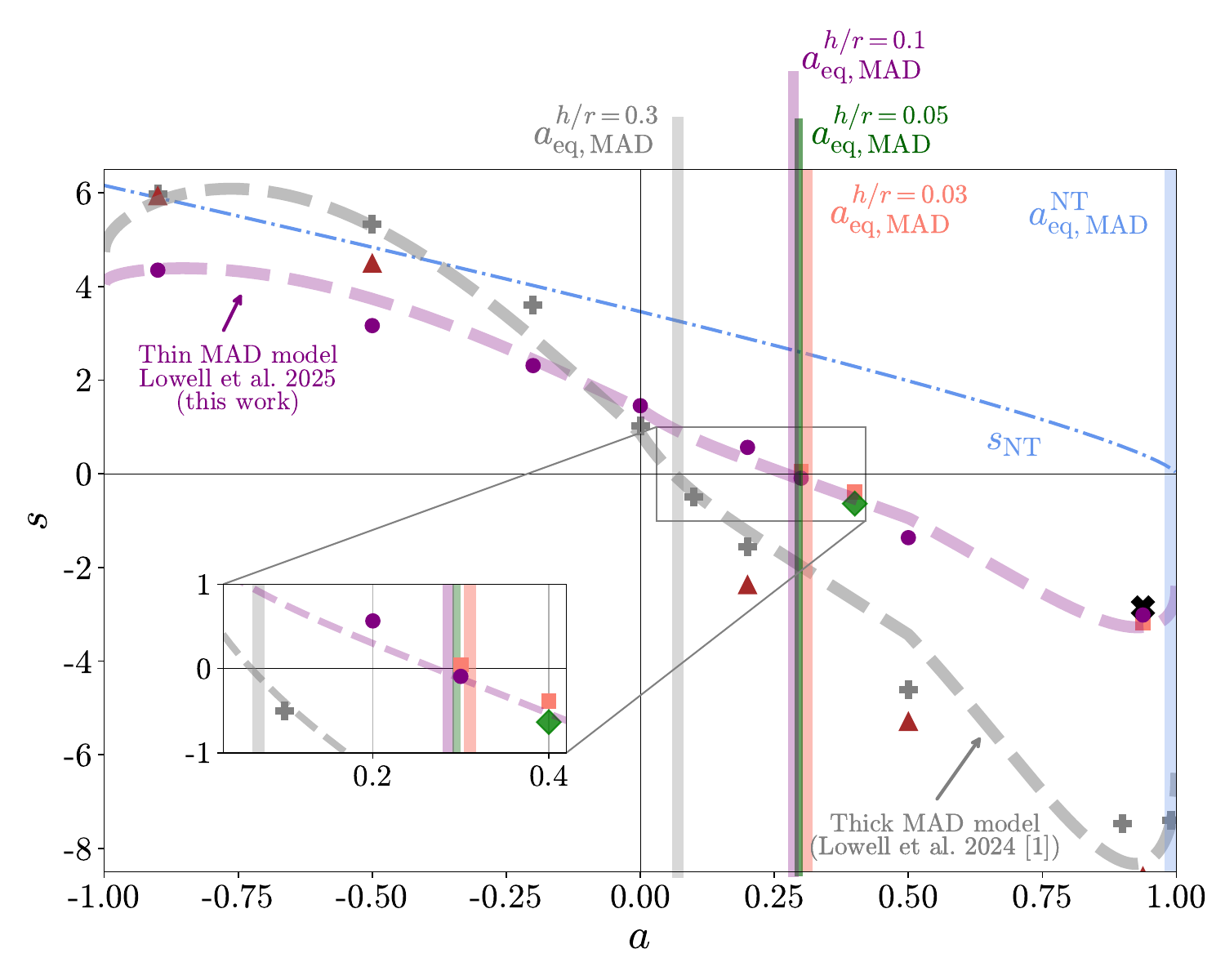}
    \caption{
     Thin MADs spin down their BHs to a low equilibrium spin, $a_\mathrm{eq,MAD}^\mathrm{thin}\approx 0.3$: although it is higher than for thick MADs, which have $a_\mathrm{eq,MAD}^\mathrm{thick}\approx 0.07$, it is still much much lower than unity. We reveal this through the plots of the spin-up parameter, $s$, vs BH spin, $a$, for several disk aspect ratios ($h/r= 0.03, 0.05, 0.1, 0.3$). The positive values of the Novikov-Thorne (NT) disk's dot-dashed blue spin-up curve indicate sustained spin up to the equilibrium spin, $a_\mathrm{eq,NT}=1$. We show the nonradiative $h/r=0.3$ MADs with $\Gamma=4/3$ with gray crosses (simulation results) and dashed gray curve (semi-analytic model of \citep{lowell_rapid_2023}). We show the $h/r=0.1$ MADs with filled purple circles, the $h/r=0.05$ MAD with a green diamond, and the $h/r=0.03$ MADs with orange squares for the radiative 2T and a black x symbol for cooled 1T simulations. We also show the higher-resolution radiatively inefficient MADs with $\Gamma=13/9$ with brown triangles. We indicate the equilibrium spin values, where the spin-up curves inferred for each of the simulation families vanish, with vertical semi-transparent lines of the corresponding colors. The values of $a_\mathrm{eq}$ for all thin MADs, $h/r \lesssim 0.1$, cluster around $a_\mathrm{eq,MAD}^\mathrm{thin} \simeq 0.3$, which is much smaller than $a_\mathrm{eq,NT} \approx 1$ and much larger than $a_\mathrm{eq,MAD}^\mathrm{thick} \approx 0.07$. 
    }
    \label{fig:spinup}
\end{figure}

Figure~\ref{fig:spinup} shows the dimensionless BH spin-up parameter, defined in Equation~\eqref{eq:s_components}, as a function of BH spin, for different disk scale heights. Each data point represents a simulation's $s$ value, which is the sum of the HD, EM, and radiation components:
\begin{equation}
    s_\mathrm{MAD} = s_\mathrm{HD} + s_\mathrm{EM} + s_\mathrm{rad}.
    \label{eq:s_components}
\end{equation}
Radiation torques are only present in the radiation transport simulations shown with orange squares. The magnitude of $s$ is an essential part of BH spin-down, as it controls the timescale over which the BH spins down to its equilibrium spin, as we discuss in Section~\ref{sec:coolmodel}.

Gray crosses in Figure~\ref{fig:spinup} show the spin-up parameter for thick, $h/r = 0.3$ MADs, and the gray dashed line represents our semi-analytic model for nonradiative MAD spin-down \citep{lowell_rapid_2023}. When \( a \sim -1 \), the BH rapidly spins up toward \( a = 0 \) due to the large spin-up parameter, \( s \). As $a$ approaches zero and \( s \) decreases, spin evolution slows until the BH reaches the spin equilibrium, \( s=0 \). For each family of simulations, Figure~\ref{fig:spinup} shows with vertical lines the equilibrium spin for each case, at which the angular momentum and energy fluxes balance: this halts further spin evolution.
Unlike NT disks, which spin up at all spins, thick MADs spin down their BHs at high spin, \( a \lesssim 1 \), where \( s < 0 \). As \( a \) decreases, \( s \) increases and eventually vanishes, at which point the BH reaches the spin equilibrium, \( a_\mathrm{eq,MAD}^\mathrm{thick} \approx 0.07 \).  

In Figure~\ref{fig:spinup}, we also present nonradiative, $h/r = 0.3$ $HR$ runs at higher resolution and $\Gamma = 13/9$ (brown triangles). These consistently yield larger $|s|$ (more negative, by $\Delta s\sim0.7{-}0.9$) values near the equilibrium spin compared to the $\Gamma = 4/3$, $h/r=0.3$ simulations (gray crosses), implying a lower equilibrium spin. This weak dependence on $\Gamma$ can account for the difference between the equilibrium spin values reported by \cite{lowell_rapid_2023} and \cite{narayan_jets_2022}.

Purple circles in Figure~\ref{fig:spinup} show thinner, $h/r = 0.1$ MADs. Similar to thick MADs, they also spin down their BHs at high spin, albeit with a factor of \( \sim 2 \) lower spin-up parameter. This reduction in the magnitude of \( s \) results in a higher equilibrium spin, \( a_\mathrm{eq,MAD}^{h/r=0.1} \simeq 0.29 \), which we mark by the vertical purple line. Although this is higher than for the thick, nonradiative MADs, it remains significantly lower than the NT disk prediction, \(a_\mathrm{eq,NT} \simeq 1\). We show with the dashed purple line our new semi-analytic model for spin-down in thin MADs (see Section~\ref{sec:coolmodel}).

Orange and green points in Figure~\ref{fig:spinup} show that as compared to \( h/r=0.1 \), \( s \) remains largely unchanged for our thinner disks, \( h/r=0.05 \) and \( h/r=0.03 \). We find only a slightly higher equilibrium spin for our thinnest ($h/r=0.03$) disks, $a_\mathrm{eq,MAD}^{h/r=0.03} \simeq 0.31$. Because for $h/r = 0.05$ we have a single simulation for $a = 0.4$, we cannot accurately compute the equilibrium spin for this thickness. However, we note that the values of the slope, $ds/da$, for the $h/r=0.03$ and $h/r=0.1$ families of simulations are consistent with each other in the region of interest, $0.2\le a \le 0.5$. Motivated by this, we adopt the slope of our $h/r=0.1$ MAD simulations to linearize the $s(a)$ dependence (illustrated with the dotted black line in Fig.~\ref{fig:res_convergence} of Appendix~\ref{appendix:resolution}) and approximately determine the value of the equilibrium spin for $h/r=0.05$: we find $a_\mathrm{eq,MAD}^{h/r=0.05}\simeq0.3$. This value is consistent with the comparable disk scale heights, falling between the curves for the thinner ($h/r=0.03$) and thicker ($h/r=0.1$) disks. The values of the equilibrium spin for our thinnest disks remain considerably lower than what is expected from razor thin NT disks.

What can cause such a difference? First, we rule out insufficient numerical resolution as a factor. In Appendix~\ref{appendix:resolution}, we present a resolution study for our thin disks with $ h/r=0.1 $ and $ h/r=0.05 $. For $ h/r=0.1 $, we find that higher resolution does not significantly affect the value of $ s $, suggesting both of the resolutions yield the same $ a_\mathrm{eq} $. This implies that in our low-resolution $h/r=0.1$ simulations (models H1a\#), the values of $ s $ are numerically converged. However, for $ h/r=0.05 $, higher resolution runs lead to significantly smaller values of $ s $ and result in $25$\% smaller $ a_\mathrm{eq,MAD}^{h/r=0.05} $ values -- reduced from $ 0.4 $ to $ 0.3 $. 

We find that a transverse resolution of \( \tilde{N}_\theta \approx 9  \) cells per thermal scale height is sufficient to accurately determine the equilibrium spin, whereas \( \tilde{N}_\theta \approx 5  \) is insufficient.

For our MAD simulation suite, achieving high resolution in the disk proper (e.g., via SMR) is essential for accurately determining the equilibrium spin for thin MADs. By comparing the torques between low- (under-resolved) and high-resolution (resolved) $h/r = 0.05$ simulations, we find that the main differences are in the HD components of both $e_\mathrm{in}$ and $l_\mathrm{in}$: under-resolving thin MADs pushes their hydrodynamics closer to the NT solution.

\begin{figure}
    \centering
    \includegraphics[width=\columnwidth]{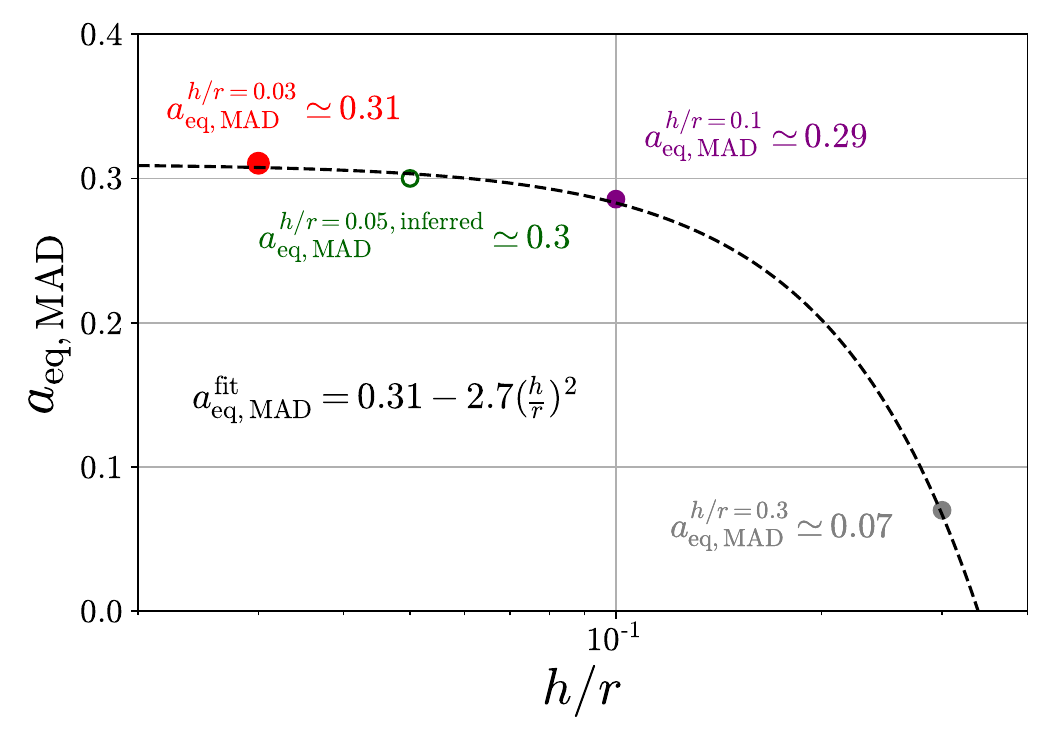}
    \caption{
     MAD equilibrium spin tends to a universal value, $a_\mathrm{eq,MAD}^\mathrm{thin} \approx 0.31$, in the limit of luminous, thin MADs, as seen in the plot of the equilibrium BH spin, $a_\mathrm{eq,MAD}$, vs disk scale height, $h/r$. Data point colors correspondingly match those in Figure~\ref{fig:spinup}. As the disk thickness decreases from $h/r=0.3$ to $h/r=0.1$, the equilibrium spin increases. However, it levels off for even thinner disks, $h/r=0.05$ and $0.03$, and approximately follows a quadratic fit, $a_\mathrm{eq,MAD}^\mathrm{fit} \approx 0.31 - 2.7(h/r)^2$, which we show with the dashed black line.}
    \label{fig:aeq_vs_hovr}
\end{figure}

Figure~\ref{fig:aeq_vs_hovr} plots the values of equilibrium spin, $a_\mathrm{eq,MAD}$, vs the disk thermal scale height, $h/r$. Each point corresponds to the equilibrium spin obtained for a suite of simulations at that $h/r$: namely, we find the $a_\mathrm{eq,MAD}(h/r)$ dependence as the root of $s(a)=0$ for a suite of simulations at that value of $h/r$. We find that as $h/r$ decreases from $0.3$ to $0.1$, the equilibrium spin significantly increases, from $a_\mathrm{eq,MAD}^{h/r=0.3}\approx0.07$ to $a_\mathrm{eq,MAD}^{h/r=0.1}\approx0.3$. Surprisingly, however, as the thermal scale height decreases further, down to $h/r = 0.05$ and $0.03$, the rise in $a_\mathrm{eq,MAD}$ slows down and appears to saturate at $a_\mathrm{eq,MAD}^\mathrm{thin}\approx0.31$. 

Figure~\ref{fig:aeq_vs_hovr} shows that a simple quadratic fit, $a_\mathrm{eq,MAD} = 0.31 - 2.7(h/r)^2$, describes the data relatively well. Because our $h/r$ coverage of thick MADs (with $h/r\gtrsim\mathrm{few}\times0.1$) has only one data point, $h/r = 0.3$, the quantitative behavior of the fit at $h/r\gtrsim0.3$ is uncertain: that $a_\mathrm{eq,MAD}^\mathrm{fit}$ becomes negative at $h/r\gtrsim0.3$ implies that the equilibrium spin becomes vanishingly small for even larger disk thicknesses. (In this context, note that the value, $h/r = 0.3$, we used to characterize thick disks is not a precise measure of their thickness, because our nonradiative simulations exhibit $h/r$ values that vary with radius, ranging from $0.2$ to $0.4$.)  

To understand this puzzling feature of MAD spin-down, below we investigate: (1) why the equilibrium spin is higher in luminous MADs compared to radiatively inefficient MADs, and (2) why the equilibrium spin converges as the disk thermal thickness decreases.

\subsection{Torques on the BH} \label{subsec:torques}

Here we unravel how the different BH torque components in our MAD simulations depend on $h/r$. We decompose the torque into the hydrodynamic and electromagnetic components, and, subsequently, into the specific fluxes that make up the spin-up parameter. 

\begin{figure}
    \centering
    \includegraphics[width=\columnwidth]{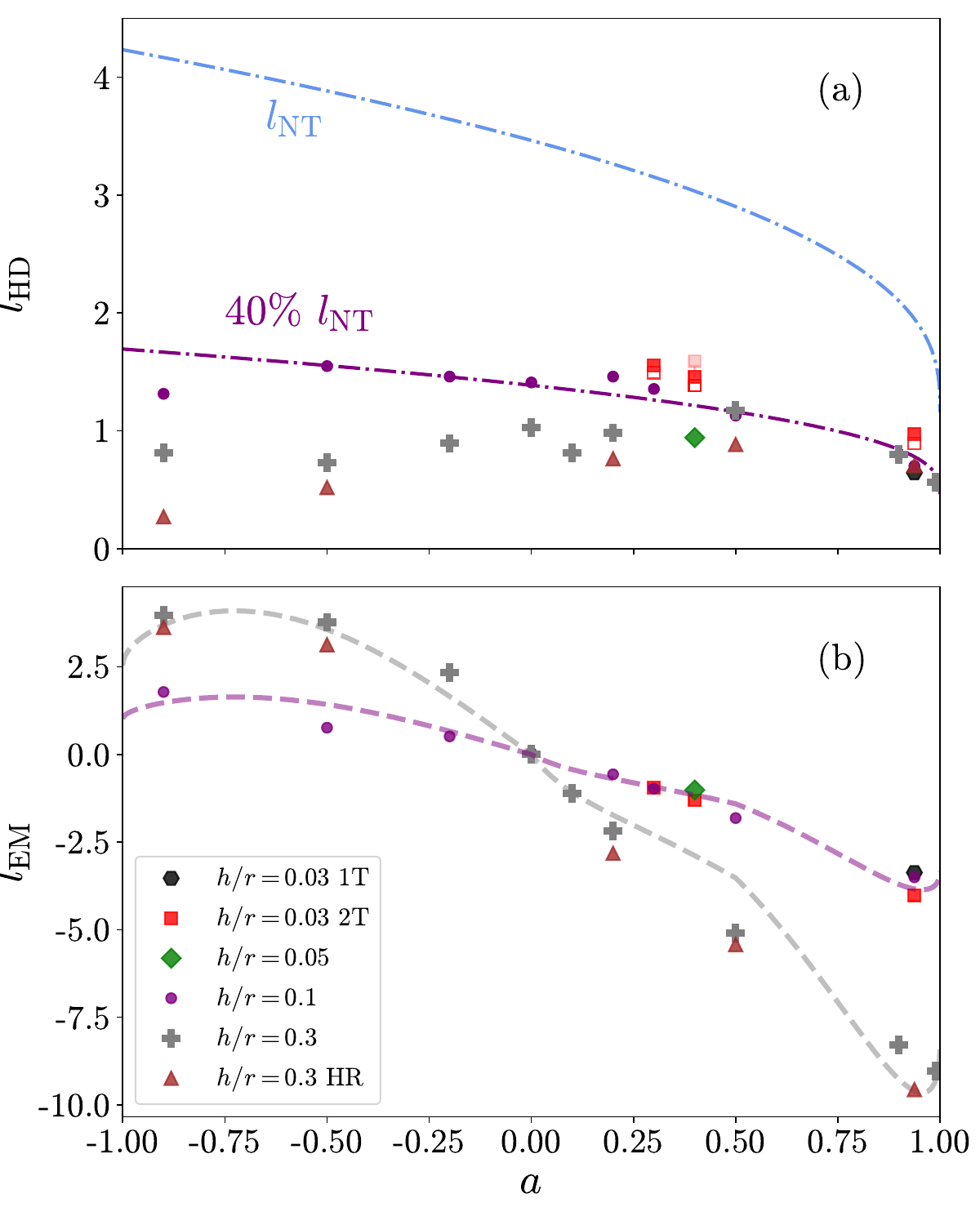}
    \caption{
    Our new analytic model (purple lines) well describes the HD and EM components of the specific angular momentum flux supplied by thin MADs ($0.03 \le h/r \le 0.1$), as seen in the plots of the specific angular momentum flux vs spin for different disk thermal scale heights: $h/r=0.03$ (2T red squares and 1T black hexagons), $h/r=0.05$ (green diamonds), $h/r=0.1$ (purple circles), $h/r=0.3$ with $\Gamma = 4/3$ (gray plus symbols), and $h/r=0.3$ with $\Gamma = 13/9$ (brown triangles). [panel~(a)]: Hydrodynamic angular momentum flux component, $l_\mathrm{HD}$, in MADs is much lower than in the NT disk (blue dashed line). Whereas $l_\mathrm{HD}$ in thick MADs (gray) remains roughly independent of spin, in thinner MADs $l_\mathrm{HD}$ decreases with increasing $a$. For $h/r=0.1$ (purple), the shapes of $l_\mathrm{HD}(a)$ and $l_\mathrm{NT}(a)$ curves are similar, so we fit $l_\mathrm{HD}(a)$ with a (purple dashed) $40\% \, l_\mathrm{NT}$ curve, which captures the data well. Interestingly, our thinnest disks ($h/r=0.03$) show very similar $l_\mathrm{HD}$ values (open red squares and filled black hexagons) to our $h/r=0.1$ results. Filled red squares show $l_\mathrm{HD} + l_\mathrm{rad}$. (We note that the $l_\mathrm{rad}$ measurement for Ra0.4, its sum with $l_\mathrm{HD}$ shown with the light filled red square, does not include a $\sigma$-cutoff to account for the numerical floors; the effects of $\sigma$-cutoffs for Ra0.3 and Ra0.94 imply that $l_\mathrm{rad}$ for Ra0.4 would be lower, shown with the dark red filled square.)
    [panel~(b)]: Electromagnetic (EM) angular momentum flux component, $l_\mathrm{EM}$, is largest in thick MADs (gray) and is lower -- and consistent -- across all thin MADs ($h/r\leq0.1$). We also show the semi-analytic EM models for $h/r=0.3$ (gray) and $h/r=0.1$ (purple dashed curves). The reduced values of $s$ for $h/r=0.3$ with $\Gamma=13/9$ (brown) relative to $\Gamma=4/3$ (gray) in Figure \ref{fig:spinup} reflect the lower contributions from both $l_\mathrm{HD}$ and $l_\mathrm{EM}$ shown here.
}
    \label{fig:lHD_lEM}
\end{figure}

Figure~\ref{fig:lHD_lEM}(a) shows the specific hydrodynamic angular momentum fluxes at the BH horizon vs BH spin for a range of $h/r$. We show the NT specific angular momentum by the blue dot-dashed line. The gray dashed curves show the specific angular momentum for the thick nonradiative disks of \citep{lowell_rapid_2023}. We show the  angular momentum flux curves for our thinner disks, with $h/r=0.03, 0.05$, and $0.1$, in red, green, and purple, respectively. 
We find that the HD angular momentum flux for thin disks is still considerably lower (by a factor $\sim2$) than the angular momentum flux of a NT disk. This is consistent with our previous findings that electromagnetic torques within the accretion disk lead to smaller than expected HD angular momentum fluxes unto the BH \citep{lowell_rapid_2023}. Even if the magnitude is different, the similarity in the spin dependence of $l_\mathrm{HD,MAD}^\mathrm{thin}$ to $l_\mathrm{NT}$ allows us to construct a simple model for this quantity, 
\begin{equation}
l_\mathrm{HD,MAD}^\mathrm{thin}=0.4 l_\mathrm{NT}. 
\label{eq:l_HD_MAD}
\end{equation}
We show this dependence as the purple dot-dashed line and describe it in Section~\ref{sec:coolmodel}. We also note that the HD angular momentum (purple dots) in thin disks by up to a factor of $\sim1.5$ larger than in thick MADs (gray crosses), especially for negative spins.

Although our thinnest, $h/r=0.03$ $1T$ and $2T$  MADs (black and red open squares, respectively), lie slightly closer to the NT curve--especially when including the angular momentum from radiation in the HD torque (red filled squares)--the difference remains small, reaching at most $25\%$ and averaging below $10\%$.

Figure~\ref{fig:lHD_lEM}(b) shows that in contrast to $l_\mathrm{HD}$, the electromagnetic angular momentum flux, $l_\mathrm{EM}$, displays a starker difference between thin and thick MADs, with a factor of $\lesssim3$ difference for high BH spin. Although the magnitudes are remarkably different, the electromagnetic angular momentum flux in thin disks follows a similar trend with $a$ as in the thick MADs.

For all disk thicknesses, the jets are the main component of angular momentum extraction out of the BH at high spin values. Thus, revealing the physical processes that set  $l_\mathrm{EM}$ is crucial to understanding the BH spin-down in MADs. Furthermore, as we have seen above, the hydrodynamic angular momentum flux does not change sufficiently strongly with disk thickness to explain the differences in $s$ between thin and thick MADs (see Fig.~\ref{fig:spinup}).

\begin{figure}
    \centering
    \includegraphics[width=\columnwidth]{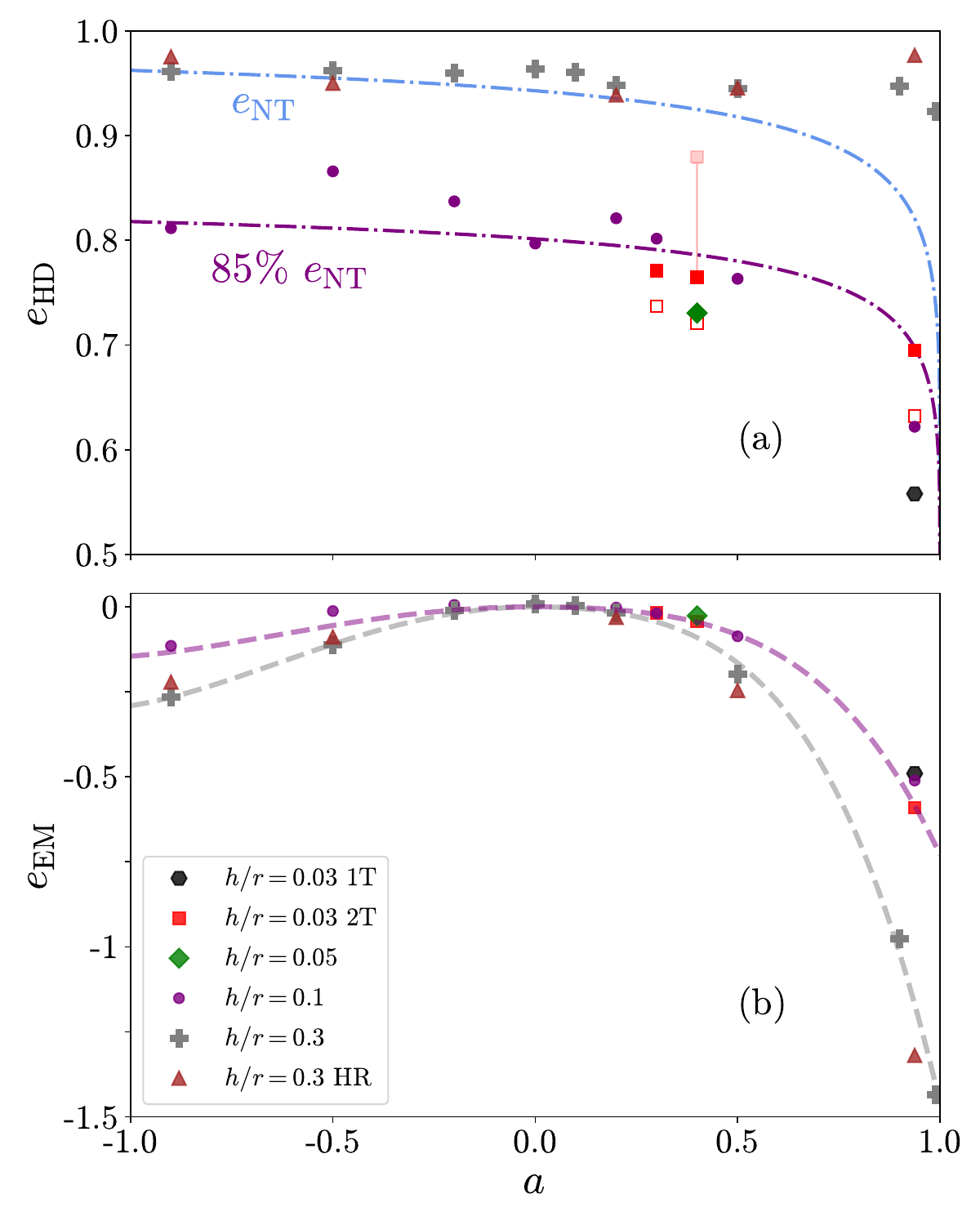}
    \caption{
    Our new analytic model (purple lines) well describes the HD and EM components of the specific energy flux supplied by all our thin MADs ($0.03 \le h/r \le 0.1$), as seen through the plots of the specific energy flux on the event horizon vs BH spin for different thermal scale heights: $h/r=0.3$ with $\Gamma=4/3$ (gray plus signs), $h/r=0.3$ with $\Gamma=13/9$ (brown triangles), $h/r=0.1$ (purple circles), $h/r=0.05$ (green diamonds), and $h/r=0.03$ (2T red squares and 1T black hexagons). [panel~(a)]: Spin dependence of HD specific energy flux, $e_\mathrm{HD}$, shows that thin MADs with $h/r=0.1$ contribute less HD energy to the BH than thick MADs with $h/r = 0.3$. The values of $e_\mathrm{HD}$ for $h/r=0.1$ appear to follow a similar shape vs $a$  as for the NT disk. We plot $85 \%$ the $e_\mathrm{NT}$ curve with the purple dashed line. The $e_\mathrm{HD}$ points for $h/r=0.03$ (red and black) follow the same trend as $e_\mathrm{HD}$ of $h/r=0.1$. Filled red squares show $e_\mathrm{HD} + e_\mathrm{rad}$ and open red squares show $e_\mathrm{HD}$. 
    (The $e_\mathrm{rad}$ measurement for Ra0.4, its sum with $e_\mathrm{HD}$ shown with the light filled red square, does not include a $\sigma$-cutoff to account for numerical floors; the effects of $\sigma$-cutoffs for Ra0.3 and Ra0.94 indicate that $e_\mathrm{rad}$ for Ra0.4 would be lower, shown with the dark filled red square). 
    We find little difference in the $e_\mathrm{HD}$ of thin $h/r=0.1$ and thinnest $h/r\leq0.03$ MAD. 
    [panel~(b)]: The dashed purple line, which well represents the EM specific energy flux of thin MADs, is simply half of the thick MAD model (gray dashed line, Eq.~\ref{etafit}). MADs extract EM energy from spinning BHs when $e_\mathrm{EM} < 0$. 
    For $h/r=0.3$, both $e_\mathrm{HD}$ and $e_\mathrm{EM}$ show little difference between the $\Gamma=4/3$ (gray) and $\Gamma=13/9$ (brown) models, indicating that the spin-up difference in Figure \ref{fig:spinup} arises from lower HD and EM angular momentum flux values (Figure \ref{fig:lHD_lEM}). Thin MADs -- all our disks with $ h/r \leq 0.1$ -- extract a similar, universal, amount of EM energy, which is approximately half of thick MADs.
    }
    \label{fig:eHD_eEM}
\end{figure}

We will return to exploring what sets $l_\mathrm{EM}$ and how it affects~$s$ below (see Eq.~\ref{eq:spinup_MAD} and Sec.~\ref{sec:jeteff_flux}). Here, we consider the other contribution to the spin-up parameter -- the energy flux (see Eq.~\ref{eq:s_param}). 
Figure~\ref{fig:eHD_eEM} shows both components of the specific energy flux on the BH horizon, $e_\mathrm{EM}$ and $e_\mathrm{HD}$. The color convention for the different thermal scale heights is the same as in previous figures. Figure~\ref{fig:eHD_eEM}(a) shows the HD component of energy. We show the NT disk expectation by the blue dashed line, which decreases as the spin increases. 
The hydrodynamic energy, $e_\mathrm{HD}$, is lower for thin than for thick MADs. However, as with $l_\mathrm{HD}$, the difference  in $e_\mathrm{HD}$ between thin and thick MADs is at most $50\%$ and on average closer to $20\%$ for most spins. The $50\%$ difference is due to the fact that, at high spin, unlike the thick MAD, whose $e_\mathrm{HD}$ stays roughly constant with spin, the thinner MAD's $e_\mathrm{HD}$ follows a trend resembling the NT disk. This allows us, similar to Eq.~\eqref{eq:l_HD_MAD}, to model thin MAD's $e_\mathrm{HD}$ as a fixed fraction of the NT hydrodynamic energy flux, 
\begin{equation}
e_\mathrm{HD,MAD}^\mathrm{thin}=0.85e_\mathrm{NT}.
\label{eq:e_HD_MAD}
\end{equation}
The thinnest, \( h/r=0.03 \) $1T$ and $2T$ MADs (black and red open and filled squares) exhibit slightly lower \( e_\mathrm{HD} \) than the \( h/r=0.1 \) MADs at moderate spin (\( a \simeq 0.1\text{--}0.5 \)). The difference is largest  at \( a = 0.9375 \), but even then it remains rather small, $\lesssim 15$\%. We conclude that our analytical approximations, eqs.~\eqref{eq:l_HD_MAD} and \eqref{eq:e_HD_MAD}, are sufficient to capture the behavior of thin MADs.

Figure~\ref{fig:eHD_eEM}(b) shows that the EM specific energy fluxes also show large differences between thin and thick MADs and at high spin exceed those of the HD contribution (Fig.~\ref{fig:eHD_eEM}a). Overall, the EM energy fluxes are about twice as high in thick as in thin MADs. This fact will help us to construct the analytic model of thin MAD BH spin-down in Sec.~\ref{sec:coolmodel}. 

The thick $h/r = 0.3$, $\Gamma = 13/9$ (brown triangles) high-resolution simulations show nearly identical values of $e_{\rm HD}$ and $e_{\rm EM}$ as the thick disk with $\Gamma = 4/3$ (gray crosses). The differences in $s$ seen in Fig.~\ref{fig:spinup} arise from the angular momentum transport: $l_{\rm HD}$ is consistently smaller in the $\Gamma = 13/9$ simulations, while $l_{\rm EM}$ is consistently larger in magnitude. Together, these effects make $s$ more negative, leading to a smaller equilibrium spin for the $\Gamma = 13/9$ case. This may explain the discrepancy between the equilibrium spins reported in \cite{lowell_rapid_2023} and \cite{narayan_jets_2022}.

The analysis above shows that jets dominate the spin-down process for medium-to-rapidly spinning BHs, $a\gtrsim0.5$.
We can decompose  the torques applied to the BH into the disk (HD) and jet (EM) contributions~\citep{moderski_black_1996}. We can then express this decomposition in terms of the spin-up parameter (Eq.~\ref{eq:s_param}; see also Eq.~20 in~\citep{lowell_rapid_2023}), 
\begin{equation}
    s_\mathrm{MAD} 
    = s_\mathrm{HD} + s_\mathrm{EM} 
    = l_\mathrm{HD} - 2a e_\mathrm{HD} - \eta_\mathrm{EM} \left( \frac{1}{k \Omega_\mathrm{H}} - 2a\right),
    \label{eq:spinup_MAD}
\end{equation}
where we have introduced the EM energy extraction efficiency, $\etaEM=-e_\mathrm{EM}$, and expressed it in terms of the EM specific angular momentum flux into the BH, $l_\mathrm{EM} = \eta_\mathrm{EM}/(k \Omega_\mathrm{H})$. Here, $k$ is the ratio of the field line angular frequency to the angular frequency of the BH horizon\footnote{We note that in Eq.~\eqref{eq:spinup_MAD} in the limit $ a \to 0 $ the first term in parentheses diverges, $/k\Omega_H\propto 1/a\to\infty$. However, because in this limit, $\eta_\mathrm{EM}\propto \Omega_\mathrm{H}^2 \propto a^2$, the apparent divergence goes away after multiplying out the factors, $\eta_\mathrm{EM}/k\Omega_\mathrm{H}\propto a \to 0 $.}, $k=\Omega_\mathrm{F} / \Omega_\mathrm{H}$.

In the remainder of Sec.~\ref{sec:results_spinevolution}, we focus on the main quantities governing the jet energy and angular momentum extraction: jet energy efficiency ($\eta_\mathrm{EM}$, Sec.~\ref{sec:jeteff_flux}) and magnetic field angular velocity ($k$, Sec.~\ref{sec:ang_vel}).

\subsection{Jet efficiency and black hole magnetic flux}
\label{sec:jeteff_flux}
\begin{figure}
    \centering
    \includegraphics[width=0.9\columnwidth]{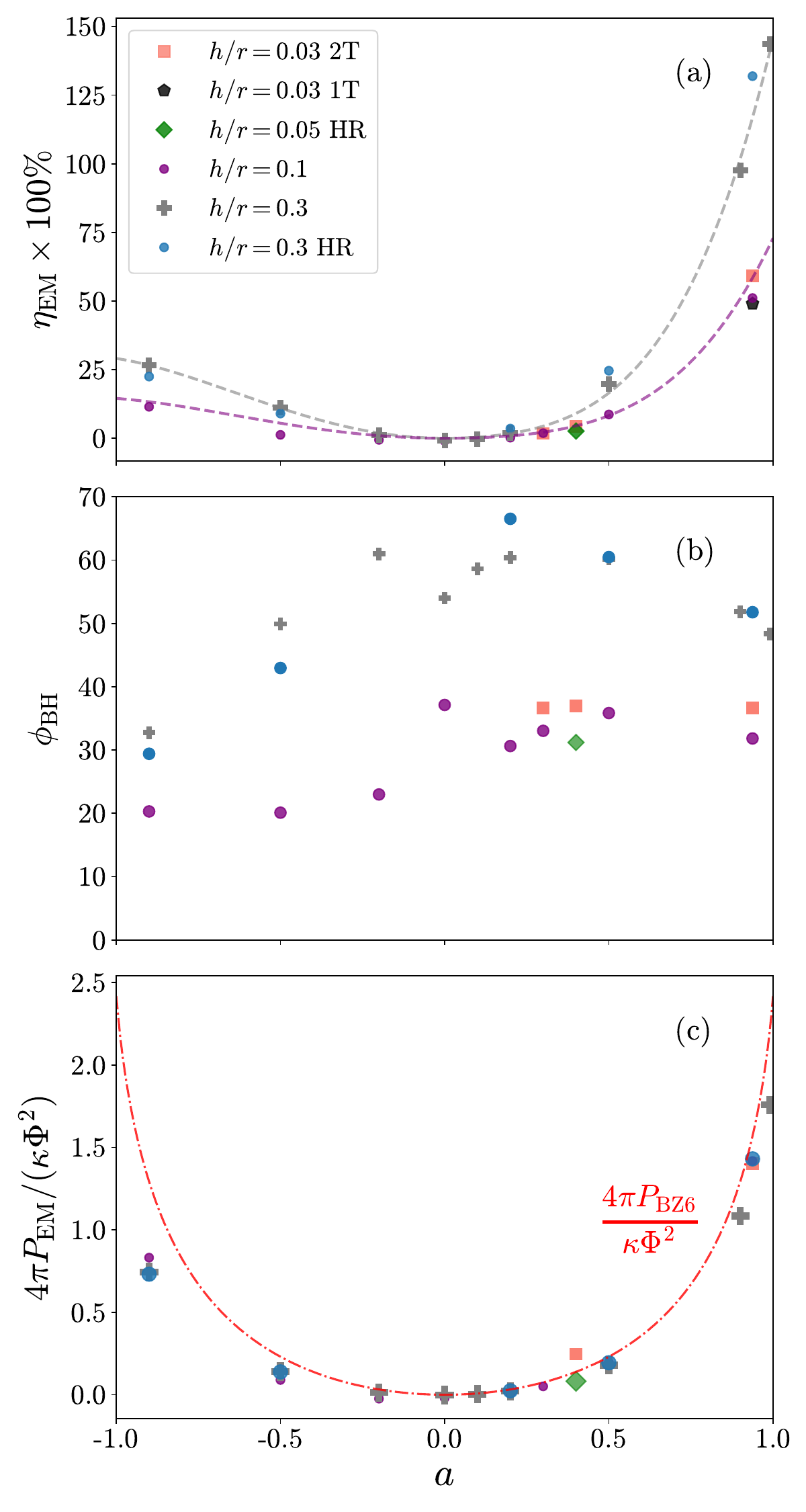}
    \caption{
    MADs of all thicknesses equally well convert BH magnetic flux into jet power. Thus, for a given mass accretion rate, MAD thickness affects the jet power only indirectly, through the BH magnetic flux. 
    [panel~(a)]: Electromagnetic (EM) efficiency vs BH spin for a range of $h/r$. At high spin, the EM efficiency of radiatively inefficient MADs ($h/r=0.3$, gray crosses and blue circles) is roughly twice that of thin MADs ($h/r=0.1$, purple circles; $h/r=0.05$, green diamonds; $h/r=0.03$, black pentagons and orange squares). Our thin MAD jet efficiency model (dashed purple line) assumes half of thick MAD jet efficiency (dashed gray line, Eq.~\ref{etafit} \citep{lowell_rapid_2023}), $\eta_\mathrm{EM,MAD}^\mathrm{thin} = 0.5\eta_\mathrm{EM,MAD}^\mathrm{thick}$, and does a good job at describing the simulation data.
    [panel~(b)]:~Magnetic flux on the BH event horizon normalized to $\sqrt{\dot{m}}$, also called the ``MADness parameter''. Thin MADs have systematically smaller BH magnetic fluxes than thick MADs.
    [panel~(c)]:~Electromagnetic (EM) power normalized to magnetic flux on the BH horizon vs BH spin for MADs of different scale heights. Although panel (a) shows that the EM energy efficiency for $h/r=0.1$ is half that for $h/r=0.3$, the magnetic flux-normalized jet powers are equivalent across all thermal scale heights (or Eddington ratios). Thus, the lower EM power in cooled disks is the result of the lower magnetic flux on the BH. The dash-dotted red line shows that the magnetic flux-normalized BZ jet power, $P_\mathrm{BZ}$, matches the simulated EM power for positive BH spins and moderately over-predicts it for negative spins.
    }
    \label{fig:eta_vs_a}
\end{figure}

We start by comparing the jet power and efficiency for different disk thermal thicknesses. Figure~\ref{fig:eta_vs_a}(a) shows EM efficiency, $\eta_\mathrm{EM}$, vs BH spin for several values of the thermal scale height, $h/r$. We find that when $h/r$ decreases from $0.3$ to $0.1$, the EM efficiency, $\eta_\mathrm{EM}$, drops by about a factor of $2$. However, when $h/r$ decreases even further, from $0.1$ to $0.03$, the $\eta_\mathrm{EM}$ decrease stalls, indicating possible convergence of $\eta_\mathrm{EM}(a)$ to a universal spin dependence, which is insensitive to $h/r$ in the limit of small disk thickness, $h/r\to 0$. Recent works also found that $\eta_\mathrm{EM}$ drops with decreasing thermal scale height \citep{scepi_magnetic_2023,dhang_energy_2024}, and we find similar values of jet efficiency to theirs for $h/r=0.1$. However, in contrast to \citep{scepi_magnetic_2023}, we find a very different jet efficiency for $h/r=0.03$, with our jets outshining theirs by a factor of $5$, although this might be due to different diagnostics for the jet efficiency.

Figure~\ref{fig:eta_vs_a}(b) shows the normalized BH magnetic flux,
\begin{equation}
  \label{eq:phibh_def}
  \phi_\mathrm{BH} = \frac{\Phi_\mathrm{BH}}{\sqrt{\dot{m} r_\mathrm{g}^2 c}},
\end{equation}
as a function of BH spin, $a$, where
\begin{equation}
    \Phi_\mathrm{BH} = \frac{1}{2} \int |B^r| dA_{\theta \varphi}
    \label{eq:Phibh_def}
\end{equation}
is the absolute BH magnetic flux, the integral is over the entire BH event horizon, and the factor of half converts it to one hemisphere. Figure~\ref{fig:eta_vs_a}(b) also shows that thicker disks have a larger normalized magnetic flux at the event horizon than thinner disks, consistent with previous findings \citep{avara_efficiency_2016,scepi_magnetic_2023,dhang_energy_2024}.

We now examine whether the weaker magnetic flux in thin MADs reduces their jet power. Via the Blandford-Znajek \citep[BZ,][]{blandford_electromagnetic_1977} mechanism, the EM jets extract the BH rotational energy at the rate \citep{tchekhovskoy_black_2010},
\begin{equation}
    P_\mathrm{BZ} = \frac{\kappa}{4 \pi c} \Phi^2_\mathrm{BH} \Omega^2_\mathrm{H}  f(\Omega_\mathrm{H}),
    \label{P_BZ}
\end{equation} 
where $\kappa$ is a constant that depends on the magnetic field geometry, $c$ is the speed of light,  $\Omega_\mathrm{H} = ac/2r_\mathrm{H}$ is the BH rotational frequency, and $f(\Omega_\mathrm{H}) = 1 + 1.38 {(\Omega_\mathrm{H} r_\mathrm{g}/c)}^2 - 9.2 {(\Omega_\mathrm{H} r_\mathrm{g}/c)}^4$ is a high-spin correction.  Here, we use the value, $\kappa=1/6\pi\approx0.053$, for a jet with a split-monopole geometry.

Equation~(\ref{P_BZ}) was derived in an idealized context of a steady state axisymmetric radial magnetic field.  How well does it describe the BH energy extraction by nonaxisymmetric collimating and time-varying magnetic fields of MADs?
To investigate this, we note that in Eq.~(\ref{P_BZ}) the BZ jet power depends on $\Phi_\mathrm{BH}$ and $a$ (via $\Omega_\mathrm{H}$). To test both of these dependencies, Figure~\ref{fig:eta_vs_a}(c) shows the EM power, $P_\mathrm{EM}\equiv\eta_\mathrm{EM}\dot{m}c^2$, normalized by $\kappa \Phi_\mathrm{BH}^2/{4\pi}$: this analytically cancels out the magnetic flux dependence and isolates the spin dependence of the EM power. 

Remarkably, once normalized this way, the EM power no longer depends on the disk scale height, $h/r$, and follows a universal spin dependence. 
In fact, the analytical spin dependence of Eq.~(\ref{P_BZ}), does a good job at explaining the simulation data for prograde MADs, but over-predicts the data for retrograde MADs by $\sim30$\% (this could be related to a larger fraction of BH magnetic flux affected by retrograde MADs).
From this, we deduce that the factor of $2$ difference in the EM efficiency between thin and thick MADs comes from the corresponding difference in $\Phi_\mathrm{BH}$. In other words, thin MADs have lower EM efficiencies than thick MADs because they have smaller $\phi_\mathrm{BH}$ values, as Figure~\ref{fig:eta_vs_a}(b) shows, and not because of deviations from the analytical expression~\eqref{P_BZ}.

We can interpret this result using Eq.~\eqref{P_BZ}. For this, 
let us we divide both sides of Eq.~(\ref{P_BZ}) by $\dot mc^2$,
\begin{align}
    \label{eq:etabz}
  \eta_\mathrm{BZ} &= \mathtextover[c]{\frac{\kappa}{4\pi}}{\frac{\kappa}{4\pi}}
                  \mathtextover[c]{\phi^2_\mathrm{BH}}{\left(\frac{\phi_\mathrm{BH}}{50}\right)^2}
                  \mathtextover[c]{\left(\frac{\Omega_\mathrm{H} r_\mathrm{g}}{c}\right)^2}{\left(\frac{\Omega_\mathrm{H} r_\mathrm{g}}{0.3c}\right)^2} 
                  f(\Omega_\mathrm{H}).\quad \text{(general MAD)}\\
\intertext{To facilitate quantitative comparison to the simulations, let us scale $\phi_\mathrm{BH}$ to the representative values, $50$ and $35$, for prograde thick and thin MADs, respectively (see Figure~\ref{fig:eta_vs_a}b) and the BH rotational rate to a typical value of the BH spin, $a = 0.9$ (which corresponds to $\Omega_\mathrm{H} r_\mathrm{g}\approx 0.3c$ and $f(\Omega_\mathrm{H})\approx1$),}
  \eta_\mathrm{BZ}   &= \mathtextover[c]{1}{\frac{\kappa}{4\pi}}
                    \mathtextover[c]{\left(\frac{\phi_\mathrm{BH}}{50}\right)^2}{\left(\frac{\phi_\mathrm{BH}}{50}\right)^2}
                     \mathtextover[c]{\left(\frac{\Omega_\mathrm{H} r_\mathrm{g}}{0.3c}\right)^2}{\left(\frac{\Omega_\mathrm{H} r_\mathrm{g}}{0.3c}\right)^2}
                     f(\Omega_\mathrm{H}), \quad \text{(thick MAD)}
  \label{eq:etabzthick}\\
  \eta_\mathrm{BZ}   &= \mathtextover[c]{\frac{1}{2}}{\frac{\kappa}{4\pi}}
                    \mathtextover[c]{\left(\frac{\phi_\mathrm{BH}}{35}\right)^2}{\left(\frac{\phi_\mathrm{BH}}{50}\right)^2}
                     \mathtextover[c]{\left(\frac{\Omega_\mathrm{H} r_\mathrm{g}}{0.3c}\right)^2}{\left(\frac{\Omega_\mathrm{H} r_\mathrm{g}}{0.3c}\right)^2}
                     f(\Omega_\mathrm{H}). \quad \text{(thin MAD)}
                  \label{eq:etabzthin}
\end{align}
Equations~\eqref{eq:etabzthick} and~\eqref{eq:etabzthin} show that these characteristic values of $\phi_\mathrm{BH}$ naturally result in $\eta_\mathrm{BZ}$ values of $1$ and $0.5$, for thick and thin MADs at $a=0.9$, respectively, in broad agreement with Fig.~\ref{fig:eta_vs_a}(a).  
Thus, we understand rather well how the spin and magnetic flux set the BH spin-down luminosity (at least for prograde MADs). However, the physics that determines the value of $\phi_\mathrm{BH}$, as a function of $h/r$, remains poorly understood.

\subsection{Angular velocity of the jet field lines}
\label{sec:ang_vel}

Now we look at the angular momentum extraction due to the EM jets. As we saw in Figure \ref{fig:lHD_lEM}, $l_\mathrm{EM}$ has a steep dependence on the thermal scale height and changes substantially between $h/r=0.3$ and $h/r=0.1$. Following \citep{moderski_black_1996,lowell_rapid_2023}, we can describe the EM specific angular momentum flux on the BH as $l_\mathrm{EM} = \eta_\mathrm{EM}/(k \Omega_\mathrm{H})$ to better understand its behavior. \citet{blandford_electromagnetic_1977} found that the optimal value of $k$ for a jet with monopolar magnetic field geometry is $k \sim 0.5$, as assumed in Eq.~\eqref{P_BZ}. Although this value maximizes energy extraction for that specific geometry, it does not necessarily maximize the efficiency for other jet geometries (e.g., parabolic). Deviations from $k \sim 0.5$ impact the efficiency of the angular momentum and energy extraction, with $k = 0$ and $k = 1$ representing the limiting cases where no energy is extracted. We therefore examine $k$ in our simulations to better understand its dependence on jet geometry, which in turn depends on disk thickness and BH spin.

\begin{figure}
    \centering
    \includegraphics[width=\columnwidth]{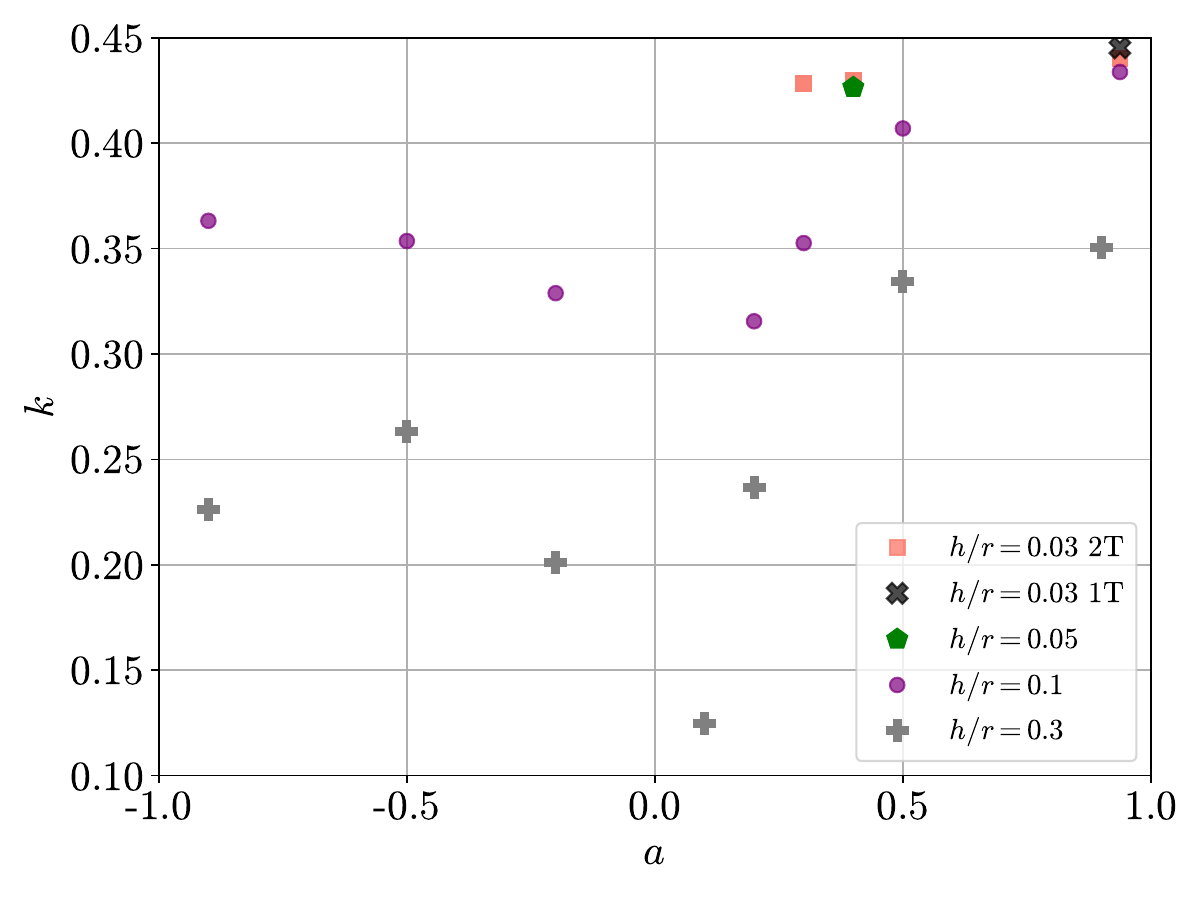}
    \caption{Thinner disks produce more rapidly rotating jets, as seen in a plot of 
    $k=\Omega_\mathrm{F}/\Omega_\mathrm{H}$ vs BH spin. The gray points show the values calculated in \citet{lowell_rapid_2023}, and the colored points show cooled disks of various scale heights (see legend). For any value of spin, thick, radiatively inefficient disks have jets with the lowest angular frequency. As the disk cools, the value of $k$ increases and appears to converge as the disk becomes thinner.}
    \label{fig:k_vs_a}
\end{figure}

In Figure \ref{fig:k_vs_a} we show the parameter $k$ as a function of spin for the different disk thermal thicknesses. We see that the thinner the disk, the larger the value of $k$. A higher $k$, indicating faster rotation of field lines, results in less efficient angular momentum extraction and thus a lower $l_\mathrm{EM}$. 
For a MAD of a given $h/r$, this leads to a higher equilibrium spin. 

The rotation velocity of the magnetic field lines, $k$, is tied to the jet structure, making it useful to examine the magnetic field lines connected to the BH in order to understand BH spin-down.
The jet structure should in turn depend on the forces acting on the disk and their balance. We examine the jet shape and force balance in Section~\ref{sec:jet_struc_force}, including the link between $k$ and the jet geometry (monopolar vs  parabolic) in section \ref{subsec:jet_struc}. But first, in Section~\ref{sec:coolmodel}, we use the dependencies of $\eta_\mathrm{EM}$ and $k$ on $a$ to construct a semi-analytical model of spin evolution for thin MADs.

\section{Semi-analytic model for BH spin-down in thin MADs} \label{sec:coolmodel}

\citet{lowell_rapid_2023} developed a physically motivated semi-analytic model that describes BH spin-down in radiatively inefficient MADs. Using our new suite of luminous MAD simulations with varying $h/r$, we now construct a model for spin-down of thinner MADs.

In panels (a) of both Figures \ref{fig:lHD_lEM} and \ref{fig:eHD_eEM},  we find that $l_\mathrm{HD}\simeq0.4 l_\mathrm{NT}$ and $e_\mathrm{HD}\simeq0.85 e_\mathrm{NT}$. Thus, we can model the hydrodynamic disk component of thin MADs as follows: $s_\mathrm{HD, thin} = l_{\rm HD} - 2a e_{\rm HD} = 0.4 l_\mathrm{NT} - 0.85 e_\mathrm{NT} \times 2a$.

The electromagnetic jet component also follows an interesting trend that is straightforward to model. As we showed in Section~\ref{sec:jeteff_flux}, $\eta_\mathrm{EM}$ for $h/r=0.1$ is approximately $1/2$ of $\eta_\mathrm{EM}$ for thick MADs. Thus, we can use the fit from \citet{lowell_rapid_2023} and model $\etaEM$ for thin MADs as $\eta_\mathrm{EM}^\mathrm{thin}=\frac{1}{2}\eta_\mathrm{EM}^\mathrm{thick}$, where
\begin{equation}
\eta_\mathrm{EM}^\mathrm{thick} \times 100 = \left\{
\begin{array}{ll}
      -19.8 a^4 + 48.9 a^2, & a < 0, \\
      106.3 a^4 + 39.5 a^2, & a \geq 0. \\
\end{array} 
\label{etafit}
\right.
\end{equation}
Figure \ref{fig:eta_vs_a}(a) shows that our above model for $\etaEM^\mathrm{thin}$ well reproduces the jet efficiency for thin MADs. 

We fit the spin dependence of the $k$ factor using the data shown in Fig.~\ref{fig:k_vs_a}, with a power law,\footnote{In this definition, we set $k(a=0)=0.2$, though strictly speaking, $k(a=0)$ is undefined. This choice, made for the quality of the fit, does not affect our results since the angular momentum extraction term goes to zero when $a=0$, as $\eta_\mathrm{EM}(a=0)=0$.}
\begin{equation}
k_\mathrm{thin} = \left\{
\begin{array}{ll}
      0.16 \,|a|^{0.15} + 0.2, & a < 0 \\
      0.24\, a^{0.4} + 0.2, & a \ge 0. \\
\end{array} 
\right.
\label{eq:kdefcool}
\end{equation}
Figure \ref{fig:lHD_lEM}, shows that our fit for $l_\mathrm{EM}=\eta_\mathrm{EM}^\mathrm{thin}/{k_\mathrm{thin}\Omega_\mathrm{H}}$, as a function of $a$, reproduces the simulation data well. 
By combining the above EM and HD components of $l$ and $e$, and replacing $\eta_\mathrm{EM}^\mathrm{thin}$ with $\eta_\mathrm{EM}^\mathrm{thick}/2$ in Eq.~\eqref{eq:spinup_MAD}, we obtain the full thin MAD semi-analytic model,
\begin{equation}
    \begin{split}
        s_\mathrm{MAD}^\mathrm{thin}  = 0.4 l_\mathrm{NT} - 0.85 e_\mathrm{NT} \times 2a  - \frac{\eta_\mathrm{EM}^\mathrm{thick}}{2} \left(\frac{1}{k_\mathrm{thin} \Omega_\mathrm{H}} - 2a\right),
    \end{split}
    \label{eq:model_thin}
\end{equation} 
for BH spin-down in thin MADs. 
Figure \ref{fig:spinup} shows our spin evolution model (Eq.~\ref{eq:model_thin}) with the purple dashed line, which reproduces the simulation data well. Although our model is tailored to $h/r=0.1$ MADs, it also well reproduces the behavior of $s$ for thinner MADs, thanks to the universality of thin MAD structure (see Section \ref{sec:forces}).

\begin{figure}
    \centering
    \includegraphics[width=\columnwidth]{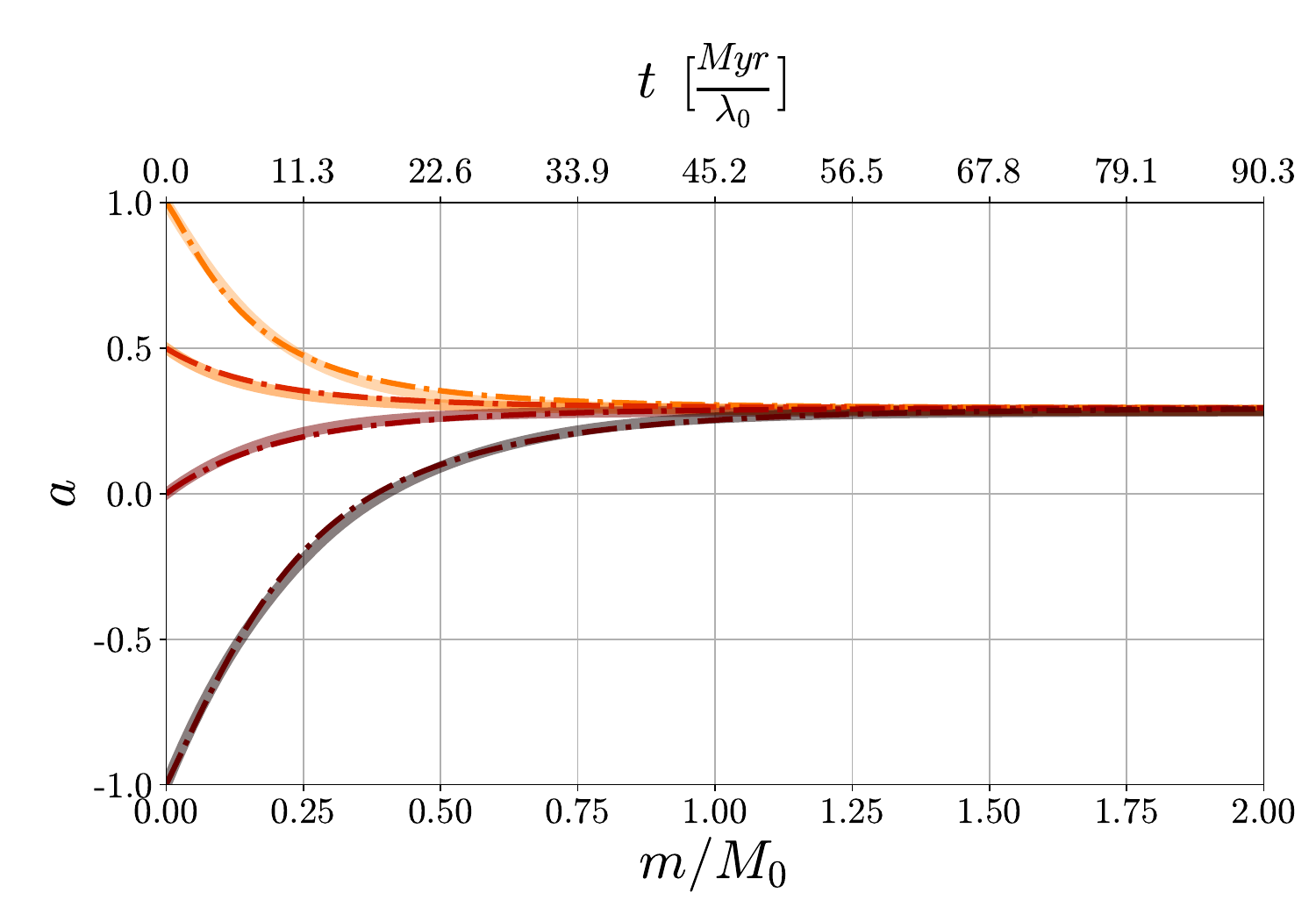}
    \caption{
    Our semi-analytic model (dashed lines, Eq.~\ref{eq:model_thin}) accurately describes BH spin evolution (solid lines) in thin MADs. We show the BH spin evolution, $a(m/M_0)$, for thin MADs with $h/r=0.1$ (models H1a\#) as a function of $m/M_0$, the ratio of accreted mass to the initial mass of the BH. When $m/M_0=1$, the BH has accreted its own (initial) mass. The upper $x$-axis gives spin evolution in the units of Myr over an initial Eddington rate, $\lambda_{0}$. We show solutions for the initial spin values, $a_0=-1$, $0$, $0.5$, and $1$: for any initial spin, the BH evolves to an equilibrium spin, $a_\mathrm{eq} \approx 0.29$. For an initial BH spin of $a=1$, a thin MAD will spin down its BH to $a=0.5$ after accreting $25 \%$ of its initial mass. For most initial spins, the BH must accrete around $85\%$ to reach equilibrium spin.
    For negative initial spin, the BH must accrete more mass than for the positive initial spins to reach $a_\mathrm{eq}$, roughly $125 \%$ of $M_0$. Dashed lines, which show the spin evolution using our semi-analytic model (Eq.~\ref{eq:model_thin}), agree well with the thin MAD spin-down model, which uses spline fits to $e_\mathrm{in}$ and $l_\mathrm{in}$.
    }
    \label{fig:a_vs_mM0}
\end{figure}

Figure \ref{fig:a_vs_mM0} shows BH spin evolution as a function of mass accreted onto the BH for thin MADs with $h/r=0.1$. For this, we  solve the coupled ODEs, Equations (\ref{eq:e_ODE}) and (\ref{eq:da_dlogM}), and use spline interpolations for the specific energy and angular momentum fluxes at the horizon, for simulations (solid lines) with $h/r=0.1$ (models H1a\#). We show solutions for the initial spin values $a_0=-1$, $0$, $0.5$, and $1$. We find that regardless of the initial spin, \( a \) evolves to an equilibrium value of \( \aeq \approx 0.3 \). Most initial spins get within 10\% of the equilibrium value after less than 85\% of the initial BH mass is accreted, while \( a_0 = -1 \) requires slightly more, roughly 100\% of the initial mass. Thin MADs with \( a = 1 \) spin down to \( a = 0.5 \) after accreting 25\% of the initial BH mass, a slower process compared to thick MADs, which only require the  accretion of 10\% of the BH mass for the same spin-down.
We also solve the ODEs using our thin MAD model (Equation \ref{eq:model_thin}, dashed lines) and confirm that our model reproduces the simulation (solid lines) well. We associate the reduced BH spin-down efficiency in thin MADs with a general decrease in angular momentum and energy extraction by the jets (see Section \ref{sec:jeteff_flux}). 

\section{Jet magnetic structure and forces}
\label{sec:jet_struc_force}

As the thermal scale height decreases in Figure \ref{fig:aeq_vs_hovr}, $a_\mathrm{eq}$ appears to converge to a universal value. This could imply that for even thinner disks, the equilibrium spin might continue to asymptotically approach a relatively small value of spin $a \lesssim 0.31$, much lower than the analytic value for NT razor-thin disks, $a=0.998$. For the thermal scale height to drop from $h/r=0.1$ to $h/r=0.03$ and leave the spin evolution unchanged, some physical property of the disk structure needs to remain independent of the scale height at $h/r \lesssim 0.1$.

In this section, we show that the magnetic structure--including the jet’s magnetic structure and the magnetic forces that squeeze the accretion disk--remains independent of scale height for $h/r \lesssim 0.1$. First, in section \ref{subsec:jet_struc}, we demonstrate that thin disks ($h/r \lesssim 0.1$) have monopolar jet morphologies, while thick disks ($h=0.3$) exhibit parabolic morphologies. Second in section \ref{sec:forces}, we show that for $h/r \lesssim 0.1$, the forces influencing disk equilibrium become predominantly magnetic. As a result, the disk structure becomes independent of the thermal scale height, leading to little or no change in the magnetic forces between models with $h/r=0.1$ and $h/r=0.03$. 

This may explain why BH spin-down by thin MADs is unaffected by the thermal $ h/r $ value: this is because the primary torque on the BH comes from the jet's magnetic fields, which appear to remain unchanged for $ h/r \leq 0.1 $.

\subsection{Jet's structure}
\label{subsec:jet_struc}
\begin{figure}
    \centering
    \includegraphics[width=\columnwidth]{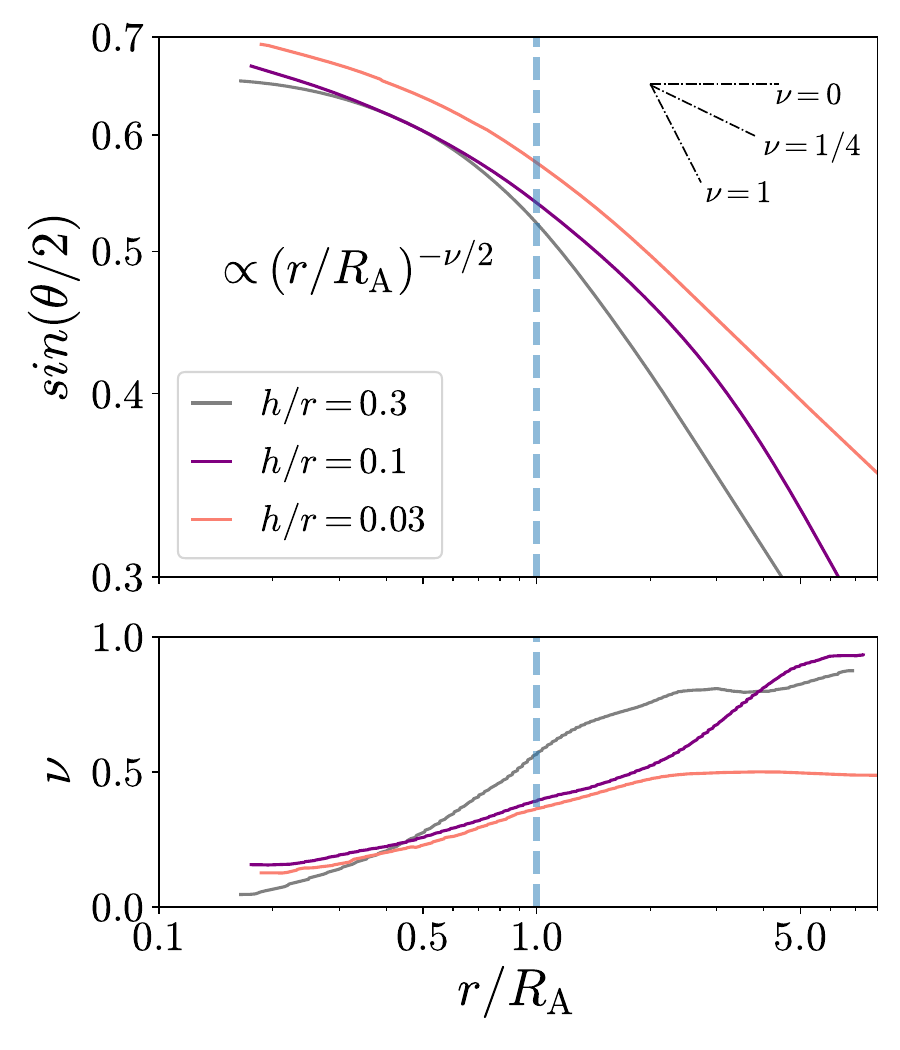}
    \caption{Thick MADs produce more collimated jets than thin MADs. Top panel: the shape of the outermost magnetic field line attached to BH with spin $a=0.9375$ for three simulations of varying scale height (here, we use the data for the $h/r=0.3$~$HR$ simulation, H3a0.94hr). We normalize the radial coordinate to the Alfv\'en radius, $R_\mathrm{A}=r_\mathrm{g}/\Omega_\mathrm{F}$, also shown by the vertical dashed blue line. For a distance greater than roughly $R_\mathrm{A}\sim 8 r_\mathrm{g}$ we find that the thinner the disk, the wider the jet opening angle. (See also Figure~\ref{fig:k_vs_a}, where thin disks have a larger value of $k=\Omega_\mathrm{F}/\Omega_\mathrm{H}$, implying that less collimated jets rotate faster than more collimated ones.)    Bottom panel: The $\nu$ jet shape parameter of the field lines in the top panel.  The jets in the thick MAD (gray) more closely follows a parabolic geometry, while jets in thin MADs (purple for $h/r=0.1$ and orange for $h/r=0.03$) are closer to a monopole geometry. This is consistent with the top panel, in which both thin disk dependencies are parallel to each other and therefore have similar $\nu$ values, whereas the thick disk has a larger $\nu$ value.}
    \label{fig:fieldlines}
\end{figure}

In this section we look at the jet's structure to see how it changes with $h/r$.
In Figure~\ref{fig:fieldlines} we plot the time- and $\varphi$-average shape of the outermost magnetic field line attached to the BH. This field line represents the jet shape at the boundary where it interacts with the wind. One can approximate the magnetic flux of a collimating field line as a power law in distance \citep{tchekhovskoy_black_2010},
\begin{equation}
    a_\varphi = r^{\nu}\left(1-\cos\theta\right),
    \label{eq:def_aphinu}
\end{equation}
where $\theta$ is the polar angle, $a_\varphi$ is the enclosed magnetic flux divided by $2\pi$ and constant along the field line, $r$ is the spherical polar distance from the center of the black hole, and $\nu$ is a parameter describing the jet shape that we generalize to depend on radius. The parameter $\nu$ sets the jet shape: $\nu=0$ corresponds to monopolar and $\nu=1$ to parabolic jets. In a realistic jet, the flux will not follow such a simple power-law. Nevertheless, we can use this approximation to understand the relationship between the jet shape and jet angular momentum flux.

From Eq.~(\ref{eq:def_aphinu}), it follows that 
\begin{equation}
    \sin \left( \frac{\theta}{2} \right) \propto  \left(\frac{r}{R_\mathrm{A}}\right)^{-\nu/2} ,
\end{equation} 
We normalize the distance to the Alfv\'en radius, $R_\mathrm{A} = 1 / \Omega_\mathrm{F}$, where the Alfv\'en velocity is equal to the outward velocity of the gas. At the Alfv\'en surface the outflowing material can no longer communicate back to the event horizon through Alfv\'en waves. Indeed, only the jet shape inside of the Alfv\'en surface can affect the jet launching and hence, the value of $k$.
Therefore, the Alfv\'en surface defines a meaningful radius where we can compare the values of $\nu$ for different simulations. 

We show jet shapes for models with $a=0.9375$ of $h/r=0.3,0.1,0.03$ using gray, purple, and orange curves, respectively. For each model, we use the value of $k$ (shown in Figure \ref{fig:k_vs_a}) to compute $R_\mathrm{A}$.  

We find that the thinner the accretion disk, the wider the jet barrel at the Alfv\'en radius. We can also see this from Figure \ref{fig:density_maps}: disks with smaller thermal thickness collimate jets less than thicker disks. The thinner the disk, the greater the jet opening angle, with magnetic field lines increasingly resembling a monopole geometry ($\nu=0$) as the disk scale height decreases. Thinner disks allow more space for jet widening, and MHD acceleration becomes most efficient with larger lever arms. As a result, field lines rotate more rapidly (with a larger $\Omega_\mathrm{F}$ and consequently greater $k$, as seen in Figure \ref{fig:k_vs_a}).

In the bottom panel of Figure \ref{fig:fieldlines} we show the $\nu$ parameter, obtained by solving for the slope in the top panel.  The value of $\nu$ at $R_\mathrm{A}$ for the thick disk is $\sim0.6$, whereas the values for the two thinner disks ($h/r=0.1$ and $h/r=0.03$) are distinctly lower, both around $\sim 0.38$.

This picture is consistent with the work of \citet{tchekhovskoy_black_2010} who studied $k$ as a function of the index $\nu$, and found that different jet geometries lead to different values of $k$. A monopole ($\nu=0$) leads to a roughly constant $k\sim0.5$, while a parabolic geometry ($\nu=1$) leads to a $k$ that decreases with angle and has an average value\footnote{We note that, in \citet{tchekhovskoy_black_2010} $k$ never dropped below $0.25$.} of $\sim 0.4$. We conclude that larger time-averaged values of $k$ are consistent with jet shapes approaching a monopole as $h/r$ decreases. We also stress that the change in $k$ and jet power is the strongest between $h/r=0.3$ and $h/r=0.1$. The change between $h/r=0.1$ and $h/r=0.03$ is almost imperceptible, which is consistent with their slopes being the same at the Alfv\'en surface.

\subsection{Decoupling of thermal and magnetic forces}
\label{sec:forces}

In this section we study the forces governing the disk equilibrium and how they change with $h/r$. We examine the thermal and magnetic forces governing disk compression and expansion in the latitudinal ($\theta$) direction, disregarding centrifugal force since it is unrelated to jet formation, BH energy and angular momentum extraction, or thermal $ h/r $.  While thermal and radiation pressures act to expand the disk, magnetic pressure plays a more complex role. Turbulent and large-scale (laminar) magnetic pressures contribute differently to the disk dynamics, influencing its overall equilibrium in distinct ways. Turbulent magnetic pressure arises from disk turbulence, providing support against collapse, whereas large-scale (laminar) magnetic pressure results from fields anchored to the jet structure, compressing the disk.
One can separate both terms using a Reynolds decomposition,

\begin{equation}
    \langle  \delta b^2 \rangle_{t,\varphi}  = \langle b^2 \rangle_{t,\varphi} - \langle b\rangle^2_{t,\varphi},
    \label{eq:reynolds_decomp}
\end{equation}
where brackets are averages in both time and $\varphi$, $\langle  \delta b^2 \rangle_{t,\varphi}$ is the turbulent magnetic pressure, $\langle b\rangle^2_{t,\varphi}$ is the large-scale (or laminar) magnetic pressure, and $\langle b^2 \rangle_{t,\varphi}$ is the total magnetic pressure.

\begin{figure}
    \centering
    \includegraphics[width=
\columnwidth]{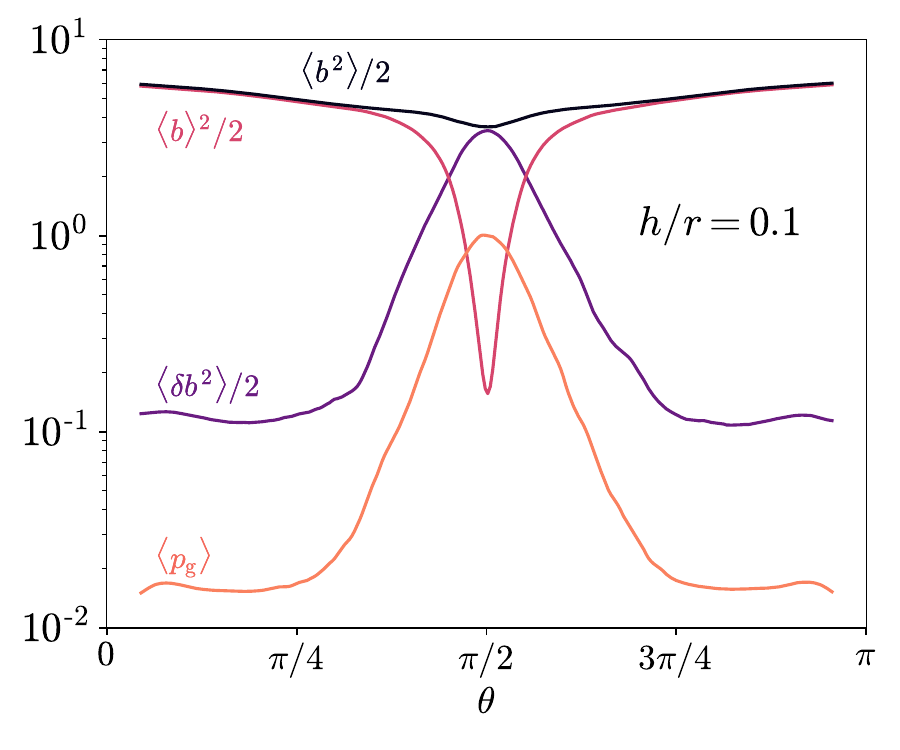}
    \caption{Vertical profiles at $r=6r_\mathrm{g}$ of time-averaged magnetic pressures for H1a0.94 normalized to thermal pressure in the midplane. Large-scale laminar magnetic pressure (magenta) dominates over turbulent magnetic pressure (purple) and thermal pressure (orange) above and below the disk. Turbulent magnetic pressure dominates the disk midplane.}
    \label{fig:vertical_pressure}
\end{figure}

We show the $\theta$-profiles of the magnetic pressures and the thermal pressure in Figure~\ref{fig:vertical_pressure} for the model H1a0.94. We see that the thermal and turbulent magnetic pressures have similar vertical profiles, whereas the large-scale laminar magnetic pressure has an inverted profile. The reason for this difference is that $\langle \delta b^2 \rangle_{t,\varphi}$ and $\langle b \rangle^2_{t,\varphi}$ play different roles \footnote{Explicit compression and expansion can be understood by examining the inverse of the gradient of the various terms in Fig.~\ref{fig:vertical_pressure}. A positive gradient, $P(\theta_2)-P(\theta_1)>0$ where $\theta_1=\pi/2$ and  $|\theta_2-\pi/2|>0$,  results in compression, while a negative gradient, $P(\theta_2)-P(\theta_1)<0$, leads to expansion.}:
(1) the laminar magnetic pressure from the jet compresses the disk and (2) the turbulent magnetic pressure, along with the thermal pressure, resists the disk compression and maintains the magnetohydrostatic equilibrium. In radiative simulations, the radiation pressure also resists disk compression, as expected.
The compressive role of $\langle b \rangle^2_{t,\varphi}$ is an established result from self-similar theory \citep{ferreira_magnetized_1995}, the role of the turbulent magnetic pressure on the vertical equilibrium is consistent with previous 3D (GR)MHD simulations \citep{salvesen_accretion_2016,jacquemin-ide_magnetic_2021,scepi_magnetic_2023}, and more recent self-similar models \citep{zimniak_influence_2024}.
Similar vertical pressure profiles are observed across all simulations at all radii, modulated by disk thickness; thicker disks exhibit broader pressure profiles. However, the relative magnitudes of the various pressures depend on disk thickness.

\begin{figure}
    \centering
    \includegraphics[width=\columnwidth]{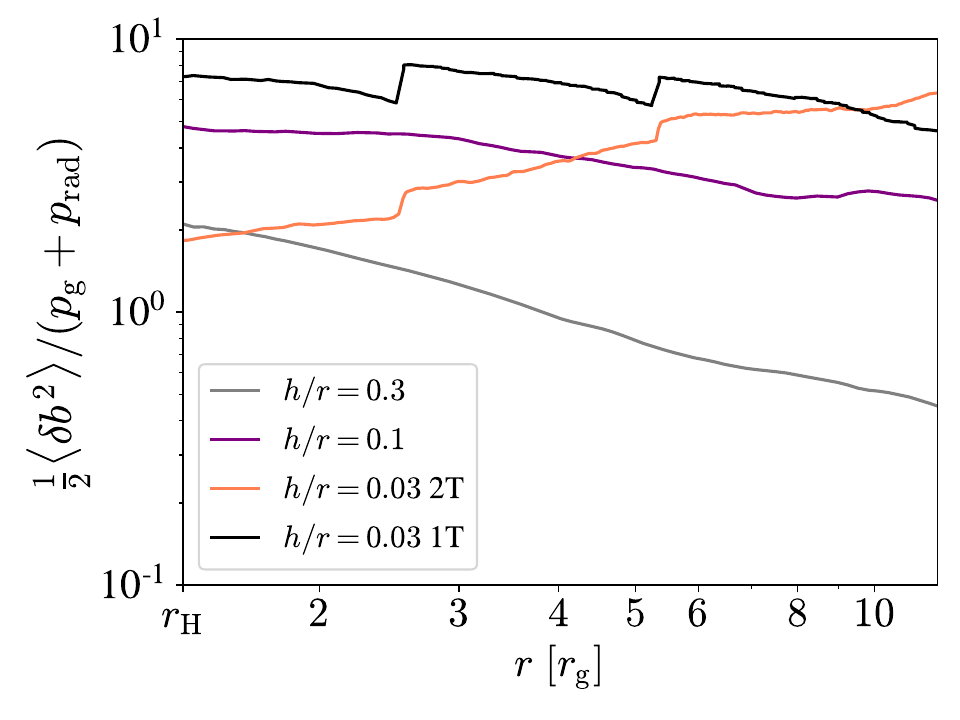}
    \caption{Ratio of turbulent magnetic pressure to thermal and radiation pressures measured in the equatorial plane for varying disk scale heights with BH spin of $a=0.9375$. The thick MAD (gray) is dominated by magnetic pressure in the inner few radii, but thermal pressure dominates beyond $4 r_\mathrm{g}$. The thin MAD (purple) is dominated by magnetic pressure overall. Magnetic pressure also dominates in Ra0.94, increasing with radius until $\frac{1}{2}\langle  \delta b^2 \rangle/(p_\mathrm{g}+p_\mathrm{rad})>6$ at $10 r_\mathrm{g}$.  We show the high-resolution simulation for $h/r=0.3$ (H3a0.94hr).
    }
    \label{fig:midplane_pressuresl}
\end{figure}

We now focus on how the relative magnitudes of the pressures depend on the disk thickness. In Figure \ref{fig:midplane_pressuresl} we plot the ratio of the turbulent magnetic pressure to the sum of the thermal and radiation pressures in the disk midplane for three simulations with $a=0.9375$ and $h/r=0.3$, $0.1$, and $0.03$, shown as gray, purple, and orange lines, respectively.  Since these pressures all act to expand the disk, this ratio indicates the dominant pressure support mechanism for the accretion disk. A clear dichotomy emerges between thick ($h/r=0.3$) and thin disks ($h/r=0.1$ and $h/r=0.03$). Thin disks are magnetically dominated at the midplane over a larger range of radii, $r\lesssim15\,r_\mathrm{g}$ compared to thick disks, which are only magnetically dominated at $r\lesssim3\,r_\mathrm{g}$. We conclude that thin disks are supported primarily by turbulent magnetic pressure, whereas thick disks are thermally supported, in agreement with previous findings \citep{scepi_magnetic_2023}. This shows that the disk thermal scale height is now decoupled from the disk vertical equilibrium and thus is not a relevant quantity for characterizing the system.

In Section \ref{subsec:jet_struc}, we demonstrated that jet shape (parabolic or monopolar) directly determines the $k$ factor, playing a key role in regulating the jet angular momentum flux from the BH. We also found that simulations with $ h/r = 0.1 $ and $ h/r = 0.03 $ exhibit identical jet shape slopes, both forming monopolar-like jets, whereas thick disks produce parabolic jets. Geometric arguments suggest a connection between jet morphology and disk thermal thickness, which we propose is mediated by the compression of the disk by the large-scale magnetic field attached to the jet.

To show this, we examine the density-normalized vertical compression force exerted by the pressure of the large-scale magnetic field, $\langle b \rangle^2_{t,\varphi}$, on the accretion disk defined as,
\begin{equation}
    F_\mathrm{b,{jet}} = -\frac{1}{ \sqrt{-g} \langle \rho \rangle_{t,\varphi}}\partial_\theta\left(\sqrt{-g}\langle b \rangle^2_{t,\varphi} /2 \right).
    \label{eq:force_b}
\end{equation}
We divide by the local density, as magnetic acceleration and compression force strength depend on density, with lower densities enabling more efficient acceleration. We ignore the curvature terms, because we found them to be negligible. We also ignore the vertical ($\theta$-) magnetic tension forces, as we also found them to be negligible \citep[this is not the case for radial tension forces, see][]{chatterjee_flux_2022}.

\begin{figure}
    \centering
    \includegraphics[width=\columnwidth] {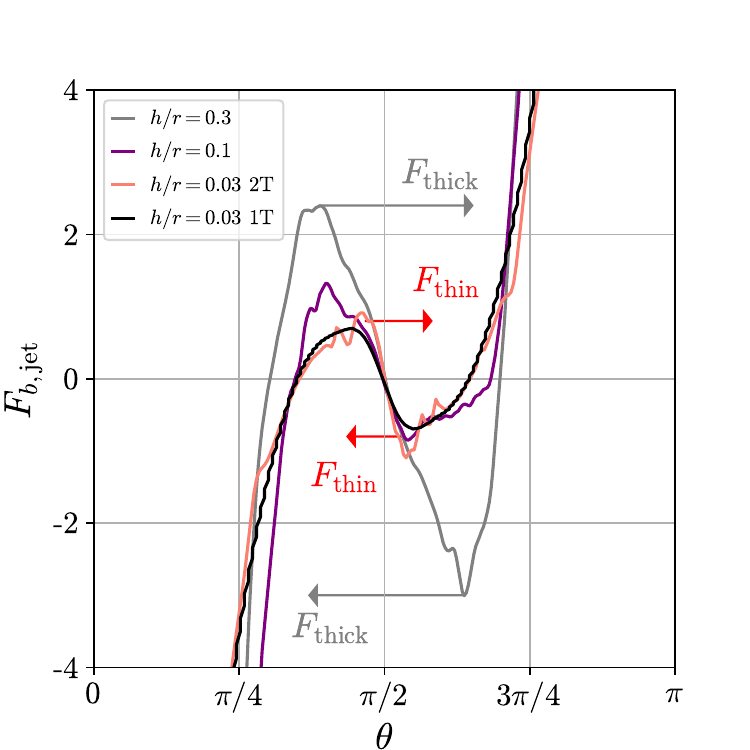}
    \caption{ All thin MADs, $0.03 \le h/r \le0.1$, experience comparable magnitude of the magnetic compression force, $F_{b,\rm jet}$, due to the twin polar jets, as shown here for BH spin, $a = 0.94$. In contrast, for thick MADs ($h/r=0.3$), the compression force is twice as strong: $F_\mathrm{thick} \sim 2 F_\mathrm{thin}$. This is consistent with thicker disks producing more powerful jets that compress the disk more. The density-normalized compression force, $F_\mathrm{b,{jet}}$, due to the large-scale magnetic field is shown as a function of $ \theta $, defined in Eq.~(\ref{eq:force_b}) and measured at $r=3\,\,r_\mathrm{g}$. Red arrows indicate the force direction, with positive forces pointing along $ \theta $ and negative forces pointing against it.  We show the high-resolution simulation for $h/r=0.3$ (H3a0.94hr).
}
    \label{fig:b_force}
\end{figure}

In Figure \ref{fig:b_force}, we present $F_\mathrm{b,{jet}}$, the density-normalized compression force measured at $r=3r_\mathrm{g}$, chosen as a representative radius close to the BH. We plot $F_\mathrm{b,{jet}}$ for four simulations with $a=0.9375$: $h/r=0.3$, $h/r=0.1$, $h/r=0.03$ 2T, and $h/r=0.03$ 1T shown as gray, purple, orange and black lines, respectively. Arrows indicate the force direction: a positive force pushes along $\theta$, while a negative force pushes in the opposite direction. All simulations exhibit similar profiles, showing compression toward the disk midplane above and below it and an expansion region above and below the compression zone. We associate the expansion region with jet expansion and confinement; however, this is not our region of interest. The magnitude of the disk compression force reveals a dichotomy between the thin ($h/r=0.1$ and $h/r=0.03$) and thick ($h/r=0.3$) disks: thicker disks display a stronger compression force above and below the disk midplane. Furthermore, all simulations of thin disks show roughly similar disk compression force magnitudes, even though their disk thicknesses differ by a factor of 3.

We find a stronger disk compression force for thicker disks, because thicker disks restrict the jet expansion, resulting in a stronger backreaction against the disk. When the disk becomes sufficiently thin ($h/r \leq 0.1$), further thinning has minimal impact since the jet occupies most of the available space. The twofold increases in $F_\mathrm{b,{jet}}$ compression for thicker vs thinner disks is consistent with the twofold increase in jet efficiency observed in Fig.~\ref{fig:eta_vs_a} for thicker vs thinner disks \footnote{The fact that $F_\mathrm{b,{jet}}$ is strongest for thick disks may seem at odds with thinner disks being more highly magnetized at the midplane (see Fig.~\ref{fig:midplane_pressuresl}). However, the ratio $\frac{1}{2}\langle  \delta b^2 \rangle/(p_\mathrm{g}+p_\mathrm{rad})$ can be misleading for comparing magnetic field strength across simulations, as thermal (and radiative) pressure also changes with $h/r$. }.

We conclude that when the disk thickness falls below a certain threshold, the disk properties are determined more strongly by the disk magnetic field structure than by the disk thermodynamics. This is evidenced by the minimal differences in $F_\mathrm{b,{jet}}$, $\eta_\mathrm{EM}$, $k$, and $\frac{1}{2}\langle  \delta b^2 \rangle/(p_\mathrm{g}+p_\mathrm{rad})$ between the simulations with $h/r=0.1$ and $h/r=0.03$. Consequently, we suggest that this is why the equilibrium spin, $a_\mathrm{eq}$, converges as the disk thickness decreases. For disk thickness values below $h/r\sim0.1$, jet properties become decoupled from the disk thermodynamics, because magnetic forces dominate over thermal forces. 

\section{Summary and Discussion} \label{sec:discussion}

\subsection{Summary}

In this study, we performed 3D GRMHD simulations of thin, luminous MADs with various disk thicknesses to investigate BH spin evolution in time. Previous works found that thick nonradiative MADs ($h/r\simeq 0.3$) result in low equilibrium spin values, $a_\mathrm{eq} \simeq 0.07$ for gas polytropic index $\Gamma=4/3$ \citep{2012JPhCS.372a2040T,2015ASSL..414...45T,lowell_rapid_2023} and $a_\mathrm{eq} \simeq 0.035$ for $\Gamma=13/9$ \citep{narayan_jets_2022,lowell_rapid_2023}. Our findings reveal, for the first time, that even in thin MADs ($h/r\leq0.1$) the BH spins down to a small equilibrium spin value, $a_\mathrm{eq}^\mathrm{thin} \simeq 0.3$. Interestingly, we observe that from $h/r = 0.1$ to $h/r = 0.03$, the difference in the equilibrium spin is negligible. This suggests that the equilibrium spin converges to a universal value $a_\mathrm{eq} \approx 0.3$. This is in stark contrast to the equilibrium spin of $a_\mathrm{eq} \approx 1$ for the analytic NT disk. Ultimately, it is straightforward to achieve a low BH spin in a system that accretes a significant amount of mass in the MAD state, be it of the luminous or nonradiative variety.
Here, we used the physical model of \citet{lowell_rapid_2023} to decompose the torques of thin MADs and analyze their BH spin evolution. As with thick MADs, the jet electromagnetic torques remain the dominant spin-down mechanism. However, we find that the higher equilibrium spin, by a factor of $4$, in thin MADs ($h/r \leq 0.1$) relative to thick MADs ($h/r=0.3$) is due to less efficient jet-driven energy and angular momentum extraction from the BH. Specifically, jets in thin MADs are half as energetically efficient as those in thick MADs. Further, thin MADs have EM angular momentum torques up to $2$ times weaker than thick MADs, for the same jet efficiency. We attribute this reduced efficiency in jet-driven angular momentum extraction to differences in jet shape between thin and thick MADs. Whereas thick MADs produce more parabolic jets, thin MADs generate comparatively more monopolar jets, which are less efficient at angular momentum extraction for the same jet energy efficiency.

We present a new semi-analytic spin-down model that accurately reproduces the spin-down timescale and equilibrium spin of thin MADs with $h/r\leq0.1$. We find that in a thin MAD, a BH with an initial spin of $a=1$ spins down to $a=0.5$ after accreting just $25\%$ of its initial mass. This is a factor of $2.5$ larger than for thick ($h/r=0.3$) MADs. Most initial spins  reach equilibrium after accreting $85\%$ of the BH’s initial mass, while $a=-1$ equilibrates after accreting $125\%$.
 
Our results indicate that some persistent MAD physical property ensures the universality of the equilibrium spin for {thermal} scale heights less than $h/r=0.1$. We propose that the explanation lies in the balance between thermal, radiation, and magnetic (large-scale or turbulent) pressures. Whereas thick MADs are thermally supported, thin MADs (of thermal $h/r \leq 0.1$) are supported by the turbulent magnetic pressure (as also found by \citet{scepi_magnetic_2023}). Furthermore, we find that in such thin MADs the vertical compression of the disk by the large-scale magnetic field pressure, associated with the jet launching, becomes independent of the disk thermal thickness. This suggests that the overall magnetic structure, which governs most of the BH spin-down, decouples from the accretion disk thermal properties for $h/r\leq0.1$. In such thin MADs, the decoupling of thermal and magnetic pressures (large-scale or turbulent) may explain why the equilibrium BH spin settles at a universal, low value, $a_\mathrm{eq} \approx 0.3$, independent of the disk thickness.

Our resolution tests (Appendix~\ref{appendix:resolution}) show that under-resolving thin MADs with \( h/r = 0.05 \) leads to a significantly higher equilibrium spin. We find that a vertical resolution of \( \tilde{N}_\theta \lesssim 5 \), where \( \tilde{N}_\theta \) is the number of cells per scale height, is insufficient to accurately determine the equilibrium spin \footnote{Difference in $\varphi$-resolution between low- and high-resolution simulation might also affect the equilibrium spin.}. This discrepancy arises from the artificially increased hydrodynamic energy and angular momentum fluxes in lower resolution runs, leading to \( \aeq \approx 0.4 \) for \( \tilde{N}_\theta \approx 5 \) and \( \aeq \approx 0.3 \) for \( \tilde{N}_\theta \approx 9 \). We emphasize that high resolution is essential for accurately modeling spin evolution in thin MADs.

It is possible that for even thinner disks, perhaps an order of magnitude thinner in $h/r$, magnetic and thermal pressures remain permanently decoupled. Alternatively, disk dynamics may still depend on $h/r$, reaching a critical value where the disk can no longer retain its magnetic flux \citep{lubow_magnetic_1995}. In that scenario, the disk would transition toward the NT state, theoretically contributing to BH spin-up. Observational evidence, such as jet quenching in the razor-thin XRB soft state, may support this idea. 
In contrast to thin MADs, which are dominated by super-thermal mainly turbulent magnetic fields at the midplane, accretion disks dominated by super-thermal toroidal fields (which is different from MADs) appear to be influenced by their thermal thickness, even if $p_\mathrm{g}$ is not dynamically relevant for the disk vertical equilibrium \citep{squire_rapid_2024}.

We compute a simple fit for the dependence of $ a_{\rm eq,MAD} $ on disk thickness, finding 
$a_{\rm eq,MAD} \simeq 0.31 - 2.7(h/r)^2.$ 
This provides a rough estimate of how $ \aeq $ might vary with the Eddington ratio, $ \lambda_\mathrm{Edd} = \dot{m}/\dot{m}_\mathrm{Edd} $, and illustrates that the equilibrium spin plateaus at $ \aeq \lesssim 0.31 $ in the limit of $h/r\to0$. In future work, we plan to extend this model across a broader range of Eddington ratios.

\subsection{Comparison with other work}

Although our work is the first to measure BH spin-down for thin MADs with $h/r\leq0.1$, previous studies have already examined jet efficiency in such disks \citep{avara_efficiency_2016,scepi_magnetic_2023,dhang_energy_2024}. For $h/r=0.1$, our results broadly agree with prior findings, showing similar jet efficiencies and magnetic fluxes at the event horizon. However, for even thinner disks ($h/r \sim 0.03$), our measured jet energy efficiencies differ significantly, with \citet{scepi_magnetic_2023} reporting jet efficiencies a factor of 5 lower. This discrepancy may stem from the differences in the measurement methods, as \citet{scepi_magnetic_2023} evaluate jet power at larger distances rather than the event horizon. Nonetheless, we note that our measurement of magnetic flux at the horizon, $\phi_\mathrm{BH}$, is consistent with \citet{scepi_magnetic_2023}, even for MADs with $h/r\sim0.03$.

Recent work by \citet{ricarte_recipes_2023} measured equilibrium spin in radiative MAD simulations but focused on thicker disks, with a minimum\footnote{We note that our measurements of $h/r$ are not entirely comparable, as they use radiative and thermal pressure to measure $h/r$.} $h/r \sim 0.18$. Direct comparison is difficult since our radiative simulations do not overlap in Eddington ratio, $\lambda_\mathrm{Edd}$, space. However, for $h/r \sim 0.18$, they report a higher equilibrium spin ($\aeq \sim 0.8$) than what we find for $h/r = 0.1$ ($\aeq \sim 0.3$). Within our model, we expect $0.07\leq\aeq\leq0.3$ for such disks, in tension with their results.

\citet{ricarte_recipes_2023} also propose a general spin-down model applicable to all Eddington ratios, which we can compare to our radiative simulations at $\lambda_\mathrm{Edd} = 0.35$ which have $h/r=0.03$ \citep{liska_formation_2022}. At this Eddington ratio they report an equilibrium spin of $\aeq \sim 1$, whereas we find $\aeq \sim 0.3$. Moreover, their results show no sign of convergence in $ a_\mathrm{eq} $ for thinner disks; instead, their equilibrium spin continues increasing as disk thickness decreases, eventually reaching values consistent with the NT disk ($ a_\mathrm{eq} = 1 $).

\subsection{Observational implications}

This work is timely for understanding BH spin measurements in x-ray binaries, where mass transfer--the process by which a donor star loses mass to a BH--is a key aspect of evolution and may also govern the BH spin evolution. Stellar-mass BH spins are directly measured via gravitational waves (GWs) in compact binary mergers \citep{abbott_binary_2019,wysocki_reconstructing_2019,ligo_scientific_collaboration_population_2023,edelman_cover_2023}, constrained in XRBs using x-ray reflection \citep{liu_precise_2008,reynolds_observational_2021,draghis_systematically_2024} and continuum fitting \citep{narayan_observational_2012,steiner_jet_2013}. However, these methods yield conflicting spin distributions: GW observations show a distribution centered around $|a|\sim 0.2$, whereas EM observations find a distribution skewed toward higher BH spins, $a\gtrsim0.5$. 
This discrepancy may arise from GW sources and XRBs belonging to distinct binary populations, which have different mass transfer histories \citep{gallegos-garcia_high-spin_2022,fishbach_apples_2022}. Spin evolution during mass transfer involves violent accretion episodes that can significantly alter the BH natal spin~\citep{qin_hypercritical_2022}. Whether these hypercritical accretion phases are linked to ultraluminous x-ray sources (ULXs) remains uncertain, but the mere fact that ULXs exist confirms that some binaries accrete far above Eddington \citep{king_ultraluminous_2023}. The spin evolution during these highly super-Eddington accretion episodes has typically been modeled using the NT or \citet{bardeen_kerr_1970} theories, which predict BH spin evolution toward high spins--consistent with the BH spin distributions observed in high-mass XRBs \citep{gallegos-garcia_high-spin_2022,qin_hypercritical_2022}. However, binaries can undergo diverse accretion spectral states, some of which launch powerful jets \citep{done_modelling_2007,king_ultraluminous_2023}. For such jetted mass transfer phases, our MAD spin evolution model can provide a more accurate description. Moreover, because it predicts lower final spins, it may yield spin distributions that align better with those measured from GW sources--potentially helping to resolve the observed discrepancy.

A leading model for long GRBs is the collapsar scenario, where a rapidly rotating massive star collapses into a BH, which launches powerful jets that produce the gamma-ray emission \citep{woosley_gamma-ray_1993}. Our previous work on thick MADs \citep{lowell_rapid_2023} applied to collapsar GRBs demonstrated that their high accretion rates and strong jets make them ideal systems for BH spin-down \citep{jacquemin-ide_collapsar_2024,wu_maximal_2024}. We showed that collapsar BHs can spin down to $\aeq = 0.07$ within seconds, well before the GRB ends.  
However, this work did not account for neutrino cooling \citep{chevalier_neutron_1989}, which alters the disk thickness and, consequently--as shown here--the equilibrium spin. Indeed, recent work by \citet{issa_black_2025} finds that due to neutrino cooling of the accreting gas, collapsar BHs spin down to a higher equilibrium spin, $\aeq \simeq 0.13$, and therefore likely fall between our cases of $h/r = 0.1$ and $h/r = 0.3$, to be consistent with our results. Their equilibrium spin measurement could serve as an initial condition for BH spin, which our model--better suited for binary evolution--could then evolve further in time.

Jet feedback is crucial for galaxy formation and the evolution of large-scale cosmological structures like galaxy clusters. Understanding BH spin evolution in luminous quasars is essential, as the changes in BH spin directly impact the energy injection into the galaxy and cluster environment. Recent quasar formation simulations, which  track accretion disk formation and evolution within a dynamically simulated galaxy, suggest the emergence of a super-thermal, toroidally dominated disk \citep{hopkins_forged_2024}. By interpolating the final snapshot of \citet{hopkins_forged_2024} simulation to \texttt{H-AMR}, \citet{kaaz_h-amr_2025} found that the central BH can accrete at super-Eddington rates, $\lambda_\mathrm{Edd}\equiv\dot{m}/\dot{m}_\mathrm{Edd}\sim 5 $. Their disk settles into a MAD-like structure similar to the thin MADs studied here. Their simplified spin-down estimates suggest spin evolution on timescales comparable to the AGN duty cycle, $\sim10\,\,\rm{Myr}$. This indicates that our results on BH spin-down are highly relevant for cosmological quasar accretion, as these systems accrete rapidly and exhibit MAD-like structures.  This is especially important if one wants to also understand the implications of AGN BH spin measurements for the BH accretion and merger history \citep{reynolds_observational_2021}.

Observations of jetted accretion disk sources (XRBs and AGN) reveal a strong correlation between the jet luminosity (\( L_\mathrm{jet} \)) and accretion luminosity (\( \dot{m} \)), independent of accretor mass, and spanning a wide range of Eddington ratios (\( \lambda_\mathrm{Edd} \)) \citep{serjeant_radio-optical_1998,corbel_radio/x-ray_2003,markoff_exploring_2003,corbel_universal_2013,zamaninasab_dynamically_2014}. This suggests little variation in jet efficiency, \( \eta_\mathrm{EM} =  L_\mathrm{jet}/\dot{m}c^2 \), across these systems. Our results show that in MADs, \( \eta_\mathrm{EM} \) remains independent of disk thickness for \( h/r \leq 0.1 \), and even when comparing thick (\( h/r = 0.3 \)) and thin (\( h/r = 0.1 \)) disks, jet efficiency decreases by only a factor of 2 for fixed $a$. Such an astrophysically small variation, within a factor of $2$, in \( \eta_\mathrm{EM} \) for fixed BH spin across different disk properties would be hard to distinguish observationally. This might explain the universal correlation between jet luminosity and accretion rate observed across diverse astrophysical systems over a wide range of Eddington ratios. 

Our results suggest that if all jetted BHs are in the MAD state, they should spin down to low spin values, \( \aeq \lesssim 0.3 \), regardless of their Eddington ratio, given sufficient time. This has significant implications for the cosmological evolution of BHs in AGN, directly impacting cosmological BH feedback. Additionally, in binary systems, these findings are crucial for LISA and LVK science, offering a potential explanation for the observed BH spin distribution.

\begin{acknowledgments}
BL acknowledges support by a National Science Foundation (NSF) Graduate Research Fellowship under Grant No. DGE-2234667. BL also acknowledges support by a Illinois Space Grant Consortium (ISGC) Graduate Fellowship supported by a National Aeronautics and Space Administration (NASA) grant awarded to the ISGC. JJ acknowledges support by the NSF AST-2009884, NASA 80NSSC21K1746 and NASA XMM-Newton  80NSSC22K0799 grants. JJ also acknowledges support from NASA 80NSSC22K0826 and NSF AST 2307983.
AT acknowledges support by NASA 
80NSSC22K0031, 
80NSSC22K0799, 
80NSSC18K0565 
and 80NSSC21K1746 
grants, and by the NSF grants 
AST-2009884, 
AST-2107839, 
AST-1815304, 
AST-1911080, 
AST-2206471, 
AST-2407475, 
AST-2510570, 
OAC-2031997. 
This work was performed in part at the Kavli Institute for Theoretical Physics (KITP) supported by grant NSF PHY-2309135.
This work was performed in part at Aspen Center for Physics, which is supported by National Science Foundation grant PHY-2210452.
This research used resources of the National Energy Research Scientific Computing Center, a DOE Office of Science User Facility supported by the Office of Science of the U.S. Department of Energy under Contract No. DE-AC02-05CH11231 using NERSC allocations m4603 (award NP-ERCAP0029085) and m2401. The computations in this work were, in part, run at facilities supported by the Scientific Computing Core at the Flatiron Institute, a division of the Simons Foundation. An award of computer time was provided by the ASCR Leadership Computing Challenge (ALCC), Innovative and Novel Computational Impact on Theory and Experiment (INCITE), and OLCF Director’s Discretionary Allocation programs under award PHY129.  This research was partially carried out using resources from Calcul Quebec (http://www.calculquebec.ca) and Compute Canada (http://www.computecanada.ca) under RAPI xsp-772-ab (PI: Daryl Haggard). This research also used HPC and visualization resources provided by
the Texas Advanced Computing Center (TACC) at The University
of Texas at Austin, which contributed to our results via the LRAC allocation AST20011 (http://www.tacc.
utexas.edu).

\end{acknowledgments}

\vspace{5mm}



\appendix

\section{Convergence test} \label{appendix:resolution}

\begin{figure}
    \centering
    \includegraphics[width=\columnwidth]{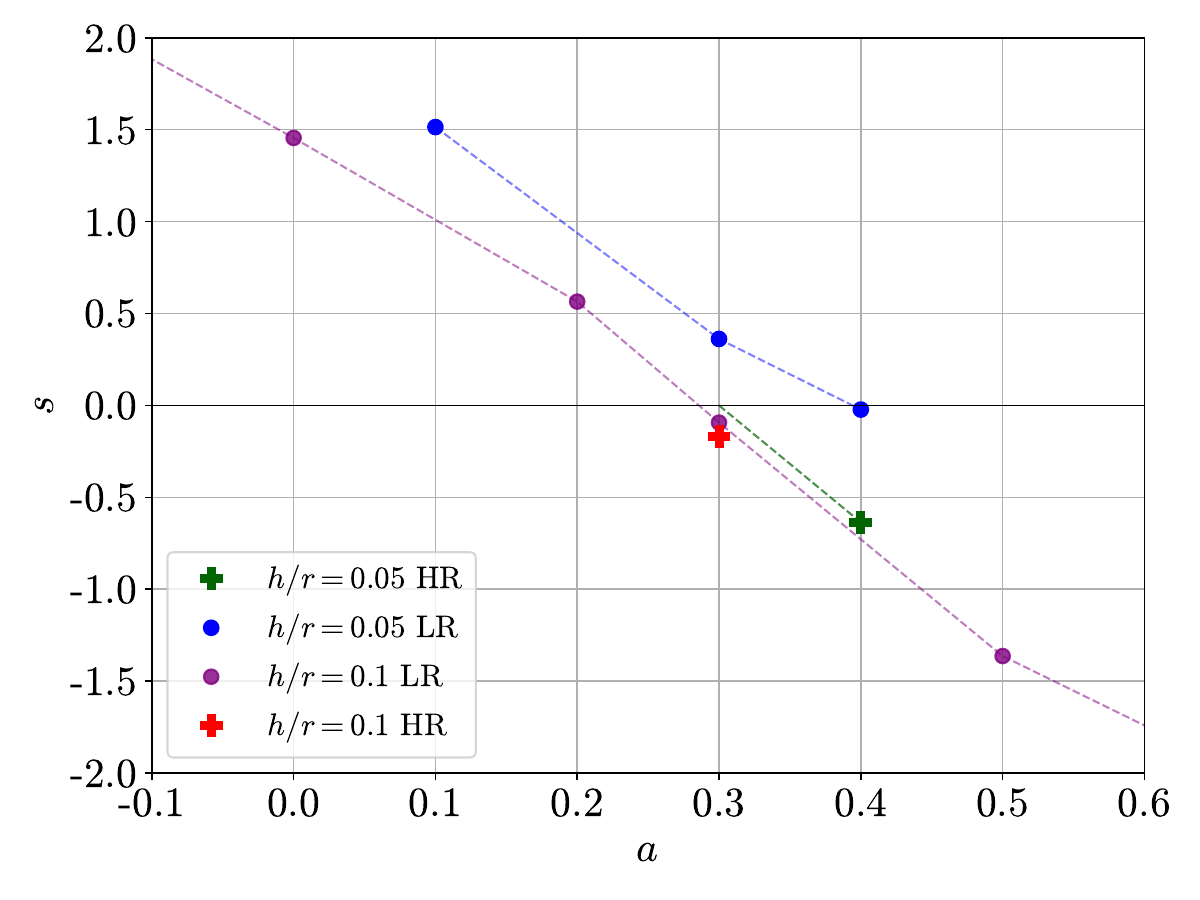}
    \caption{Thin MAD simulations need a vertical resolution of $\tilde N_\theta\gtrsim 9$ cells per scale height to yield accurate values of the spin-up parameter, as seen in a convergence study for thin MADs (with $h/r=0.1$ and $0.05$). Purple and blue circles show the low-resolution (LR) simulations and red and green crosses show their high-resolution (HR) counterparts (see Table~\ref{tab:sim_dets_hovr1}). The figure only shows results for spins between $-0.1$ and $0.6$ to highlight the convergence near the equilibrium spins.}
    \label{fig:res_convergence}
\end{figure}

We compare low- and high-resolution simulations for MAD scale heights of $h/r=0.05$ and $h/r=0.1$. Figure~\ref{fig:res_convergence} shows in purple the spin-up parameter values for our fiducial $h/r=0.1$ simulation suite: it uses the standard (``low'') resolution and  yields an equilibrium spin of $a_{\rm eq,MAD}^{h/r=0.1} = 0.29$. To verify that these results are converged with the numerical resolution, we ran a higher-resolution simulation on OLCF \emph{Summit} supercomputer and show its result in red. The consistency of $s$ between the resolutions demonstrates that the lower resolution simulations are sufficient to accurately determine the $s$ values for $h/r=0.1$ MADs and are therefore used for the analysis in the main text.

For thinner, $h/r=0.05$ MADs, Figure~\ref{fig:res_convergence} reveals significant differences in $s$ between low-resolution simulations (shown in green and sufficiently small to be run on a workstation) and high-resolution simulations (shown in blue and run on OLCF \emph{Summit}). In fact, low-resolution simulations systematically overestimate the spin-up parameter, $s$ (by $\Delta s \sim 1$), and the equilibrium spin, $a$ (by $\Delta a\sim 0.1$). Analyzing individual contributions to $s$ (not shown) reveals that this discrepancy arises from an overestimation of hydrodynamic energy, $e_{\rm HD}$, and angular momentum, $l_{\rm HD}$, fluxes. This may be due to our low-resolution $h/r=0.05$ simulations not fully resolving the fastest-growing MRI mode, reducing turbulent dissipation and impacting these fluxes.

Summing up, the resolution of $\tilde N_\theta = 5$ ($5$ cells per disk scale height) is insufficient to resolve thin MADs (with $0.05 \lesssim h/r \lesssim 0.1$). To recover the accurate values of $s$, we find that we need $\tilde N_\theta \simeq 10$. 
For this reason, we do not use the low-resolution $h/r=0.05$ simulations for the analysis in the main text.

\bibliography{Cooleddiskspindown}

\end{document}